\documentclass[authoryear, preprint, 12pt]{elsarticle}

\usepackage{amsmath, amssymb}   % Math symbols
\usepackage{graphicx, subcaption} % Figures and subfigures
\usepackage{listings}           % Code formatting
\usepackage{tabularx, booktabs} % Tables
\usepackage{url}                % URLs
\usepackage{xcolor}             % Color formatting
\usepackage{natbib}
\usepackage[margin=1in]{geometry} 
\usepackage{hyperref}

\makeatletter
\def\ps@pprintTitle{%
  \let\@oddhead\@empty
  \let\@evenhead\@empty
  \def\@oddfoot{\footnotesize\itshape Preprint}%
  \let\@evenfoot\@oddfoot
}
\makeatother

\title{Gabor-Enhanced Physics-Informed Neural Networks for Fast Simulations of Acoustic Wavefields}

\author{
	\normalsize Mohammad Mahdi Abedi\textsuperscript{1,*},  
	David Pardo\textsuperscript{2,1,3},  
	Tariq Alkhalifah\textsuperscript{4} \\  
	\footnotesize  
	\textsuperscript{1}Basque Center for Applied Mathematics, Bilbao, Spain \\  
	\footnotesize  
	\textsuperscript{2}University of the Basque Country, Department of Mathematics, Spain \\  
	\footnotesize  
	\textsuperscript{3}Ikerbasque, Basque Foundation for Science, Bilbao, Spain \\  
	\footnotesize  
	\textsuperscript{4}King Abdullah University of Science and Technology, Thuwal 23955-6900, Saudi Arabia \\  
	\footnotesize  
	\textit{Emails:} mabedi@bcamath.org, david.pardo@ehu.es, tariq.alkhalifah@kaust.edu.sa  
}

\cortext[cor1]{Corresponding author}

\begin{document}
		% Set figure path
		\graphicspath{{figures/}}
	
	% Front Matter
	\begin{frontmatter}
		
		\begin{abstract}
Physics-Informed Neural Networks (PINNs) have gained attention for solving partial differential equations, including the scattered Helmholtz equation, due to their flexibility and mesh-free formulation. However, their performance suffers from low-frequency bias, particularly in high-frequency wavefield simulations, limiting convergence speed and accuracy. To address this, we propose a novel and simplified PINN framework that incorporates explicit, trainable Gabor basis functions to efficiently capture the localized and oscillatory nature of wavefields. Unlike previous Gabor-based PINNs that rely on multiplicative filters or auxiliary networks to learn Gabor parameters, our approach redefines the network's task as learning a nonlinear mapping from input coordinates to a custom Gabor coordinate system, where a Gabor function captures the dominant oscillatory behavior of the wavefield. This formulation absorbs the effect of two Gabor parameters into the learned mapping, reducing computational complexity and eliminating the need for manual tuning of hyperparameters. We also present an efficient formulation for incorporating a Perfectly Matched Layer (PML) into the training by deriving real-valued loss components and introducing an analytical expression for the background wavefield. Numerical experiments on various velocity models show that our Gabor-PINN achieves faster convergence, higher accuracy, and greater robustness to architectural design and initialization compared to both traditional PINNs and prior Gabor-based methods. The improvement lies not in adding architectural complexity---as is common in enhanced PINNs---but in absorbing this complexity into the learned coordinate transformation, making the method both simpler and more effective. Our implementation is publicly available to support reproducibility and future research.
		\end{abstract}
		
		
	\end{frontmatter}

\section{Introduction}

Wavefield simulations governed by the Helmholtz equation, the frequency-domain form of the wave equation, are critical for understanding wave propagation phenomena in geophysics and acoustics. The Helmholtz equation is particularly relevant for seismic applications such as full-waveform inversion (FWI) and imaging \citep{pratt1999, sirgue2004}, as well as for emerging fields like offshore wind farm site characterization, where accurate subsurface imaging is essential for foundation design and geohazard assessment.

 In many practical scenarios---especially those involving complex subsurface structures---it is common to reformulate the Helmholtz equation in terms of the scattered wavefield to isolate perturbations caused by medium heterogeneities and avoid the singularity introduced by the point source in the solution \citep{Alkhalifah2021}. However, solving it numerically poses significant challenges due to its highly oscillatory solutions, which require high-resolution grids, leading to substantial computational costs \citep{Virieux1986, Marfurt1984}. Applications in wind farm site characterization typically require higher frequencies to resolve fine-scale near-surface features, further increasing computational demands. Traditional numerical methods suffer from issues such as numerical dispersion and increasing computational complexity in high-dimensional settings \citep{Hesthaven2007, Ainsworth2004}. Although dimensionality reduction offers more efficient computations, the inversion of the stiffness matrix associated with the Helmholtz equation remains computationally prohibitive, particularly for high-frequency applications or complex velocity models \citep{wu2023}.

Physics-Informed Neural Networks (PINNs) have demonstrated the ability to learn functional solutions to partial differential equations (PDEs) \citep{raissi2019}, including the Helmholtz equation \citep{song2021}. PINNs learn a continuous function that satisfies the governing conditions---which include the PDE, boundary conditions, and often initial or physical constraints---are incorporated into the loss function, guiding the training process without the need for labeled data. This enables unsupervised training and flexible domain representation \citep{llanas2006}. They are advantageous for irregular geometries, high-dimensional problems, and can also solve PDEs with problem-dependent parameters \citep{lu2021learning,zou2024seismic,baharlouei2025least}, making them adaptable for a wide range of applications.

However, PINNs face limitations in training cost and accuracy for complex functions \citep{wu2018a, song2023}. High training costs stem from the inherent low-frequency bias of neural networks, making high-frequency solutions challenging to achieve \citep{neal2019, wang2021eigenvector, Alkhalifah2024}.

To address these challenges and improve the modeling of highly oscillatory solutions, we enhance the standard PINN framework by incorporating domain-specific basis functions tailored to represent wave phenomena. Specifically, we propose a Physics-Informed Neural Network augmented with simplified Gabor basis functions. Gabor functions \citep{gabor1946}, known for their ability to represent localized oscillatory features, are well-suited for wavefield modeling. In our approach, the network learns a nonlinear transformation of the spatial inputs to a custom Gabor coordinate system, which is then used in the penultimate layer to modulate the output through a simplified Gabor function. This redefinition of the network's task enables the learned function to become smoother, as the oscillatory behavior of the wavefield is captured by the Gabor functions.

Our formulation introduces a compact representation in which a learned displacement parameter, in conjunction with the Gaussian envelope, encapsulates both the magnitude and center of the original Gabor functions. This simplification reduces the number of hyperparameters, trainable parameters, and overall computational complexity.

Additionally, we incorporate Perfectly Matched Layers (PMLs) into the training domain to improve the scattered wavefield simulation near domain boundaries. To avoid complex-valued training, we explicitly compute the real and imaginary parts of the loss function for the scattered Helmholtz equation with PML. Furthermore, we propose an  analytical expression for the background wavefield that is exact within the physical domain and serves as an approximation within the PML region. These modifications allow for efficient and stable inclusion of PMLs during training.

The rest of this paper is organized as follows. Section \ref{sec:relatedwork}  reviews related work, including recent advances in PINNs and prior Gabor-based approaches. Section \ref{sec:mathematical_formulation} introduces the mathematical formulation of the scattered Helmholtz problem and the incorporation of a Perfectly Matched Layer (PML). In Section \ref{sec:methodology}, we define the loss function and introduce our trainable Gabor functions. Section \ref{sec:implementation} details the implementation of our proposed Gabor-PINN method and its weight initialization. Section \ref{sec:numerical_results} presents numerical experiments across various velocity models and frequencies, including comparisons with other PINN variants, robustness to initialization, and evaluation of the PML effect. Finally, we summarize our findings and discuss potential directions for future work.

\section{Related work}
\label{sec:relatedwork}
Recent advancements in Physics-Informed Neural Networks (PINNs) have aimed to overcome core limitations in modeling high-frequency solutions. Feature expansion techniques, such as sinusoidal and Fourier feature embeddings \citep{tancik2020fourier}, help mitigate the spectral bias of standard multilayer perceptrons, enabling better representation of oscillatory functions \citep{huang2024, chai2024overcoming}. Multiscale and domain decomposition approaches \citep{jagtap2020extended, wang2021eigenvector} improve scalability and convergence by dividing the problem into subdomains or frequency bands. Adaptive activation functions \citep{jagtap2020locally, alsafwan2021, waheed2022kronecker}, including frequency-modulated variants \citep{sitzmann2020implicit}, dynamically increase representational capacity. Novel architectures, like Physics-Informed Kolmogorov–Arnold Networks \citep{shukla2024comprehensive} and generative PINNs \citep{chai2024overcoming, taufik2025latentpinns}, offer improved expressivity for complex PDEs, while enhanced gradient-based optimization strategies \citep{li2022gradient} further stabilize training. A recent survey by \citet{toscano2025pinns} provides a comprehensive overview of these developments.

These advances set the stage for our proposed framework, which addresses high-frequency wavefield modeling through domain-specific basis functions. Among such basis functions, Gabor functions have been shown to effectively capture oscillatory features inherent in wavefields. \cite{huang2023b} uses Gabor functions as multiplicative filters within a PINN architecture. They define each hidden layer's output as an element-wise multiplication of the linear projection of the previous layer's output and a Gabor filter function directly applied to the input coordinates. This architecture, known as a multiplicative filtered network (MFN), aims to embed known wavefield characteristics, such as frequency, into the training process to achieve faster convergence. They realized that proper initialization of the Gabor function's wavelength parameter is crucial to its stability and to avoid issues like lack of smoothness in the represented function and its gradients. They link this hyperparameter to wavefield frequency and show that defining a proper scaling factor is essential for convergence, particularly at higher frequencies. However, this architecture increases the computational complexity due to the repeated application of Gabor filters at input and all hidden layer, which introduces more function evaluations, making the network more expensive to train.

\cite{Alkhalifah2024} extended this concept by using Gabor functions as explicit basis functions in the penultimate layer of the neural network. Here, \textit{explicit} means that the Gabor functions are directly incorporated into the architecture, rather than being implicitly learned through standard network weights. In their approach, the magnitude and centers of these Gabor basis functions are not fixed but are learned by two neural networks. They demonstrated that this explicit representation significantly improves the ability of PINNs to model oscillatory wavefields, leading to faster convergence and higher accuracy compared to traditional PINNs. However, training multiple coupled networks jointly can introduce instabilities due to interdependent optimization dynamics, potentially resulting in slower or less reliable convergence. Specifically, the performance and final accuracy of their method highly depend on a proper selection of hyperparameters, including the number of neurons in their auxiliary network and the overall architecture, which often requires a trial-and-error tuning for optimal results. 

In Sections \ref{sec:methodology} and \ref{sec:implementation}, we propose a more compact incporporation of Gabor functions into the PINN framework, and in Section \ref{sec:comparison}, we present numerical comparisons that evaluate the differences between our proposed method and existing implimentations of Gabor functions in PINNs.

	\section{Mathematical Formulation}
	\label{sec:mathematical_formulation}
	
	\subsection{Helmholtz Equation for Wavefield Simulation}
	
	The wavefield simulation in this study focuses on solving the Helmholtz equation, which represents the wave equation in the frequency domain. For an isotropic medium with wave velocity \( v(\mathbf{x}) \), the Helmholtz partial differential equation (PDE) is given by:
	\begin{equation}
		\left(\nabla^2 + \frac{\omega^2}{v^2(\mathbf{x})}\right) u(\mathbf{x}) = f(\mathbf{x}),
		\label{eq:helmholtz}
	\end{equation}
	where \( \nabla^2 \) is the Laplacian operator, \( \omega \) is the angular frequency, and \( f(\mathbf{x}) \) represents the source function.
	
	The solution $u(\mathbf{x}) = u^r(\mathbf{x}) + \mathrm{i} u^i(\mathbf{x})$  is a complex-valued wavefield, where \( \mathbf{x} = \{x, z\} \) denotes the spatial coordinates in a 2D domain. Since $f(\mathbf{x})$ is modeled as a point source (Dirac's delta function), this generates a point singularity in the solution. To address this, the wavefield $u(\mathbf{x})$ is decomposed into a background wavefield $u_0(\mathbf{x})$ and a scattered wavefield $u_s(\mathbf{x}) = u(\mathbf{x}) - u_0(\mathbf{x})$. The background wavefield \( u_0(\mathbf{x}) \) is analytically calculated in a homogeneous medium with a constant background velocity \( v_0 \), defined as the velocity at the source location \( \mathbf{x}_f \), i.e., \( v_0 = v(\mathbf{x}_f) \). The analytical background wavefield \( u_0(\mathbf{x}) \) satisfies the Sommerfeld radiation condition.
	
	Substituting this decomposition into equation~\ref {eq:helmholtz} yields the Helmholtz PDE for the scattered wavefield:
	\begin{equation}
	\left(\nabla^2 + \frac{\omega^2}{v^2(\mathbf{x})}\right) u_s(\mathbf{x}) = -\omega^2 \delta m(\mathbf{x}) u_0(\mathbf{x}),
\end{equation}
where $\delta m(\mathbf{x}) = \frac{1}{v^2(\mathbf{x})} - \frac{1}{v_0^2}$ represents the perturbation in slowness squared.

\subsection{Perfectly Matched Layer}  
Perfectly matched layers (PMLs) are a widely used approach for simulating wave propagation in unbounded domains by surrounding the computational domain with artificial absorbing layers. These layers effectively attenuate outgoing waves, allowing for truncation of the computational domain without generating reflections at the boundaries. The PML technique achieves this by modifying the wave equation through complex coordinate stretching, which introduces anisotropic damping within the layer \citep{berenger1994perfectly}.

The scattered Helmholtz equation with PML reads (see \ref{app:PML} for derivation):  
\begin{equation}
	\frac{\partial}{\partial x} \left( \frac{e_z}{e_x} \frac{\partial u_s}{\partial x} \right) +
	\frac{\partial}{\partial z} \left( \frac{e_x}{e_z} \frac{\partial u_s}{\partial z} \right) +
	e_x e_z \omega^2 m(x) u_s = - e_x e_z \omega^2 \delta m(x) u_0(x),
	\label{eq:scattered_helmholtz_PML}
\end{equation}
where the complex stretching factors \( e_x \) and \( e_z \) are defined as  
\begin{equation}
	\begin{aligned}
		e_x &= 1 - i c l_x^2, \\
		e_z &= 1 - i c l_z^2,
	\end{aligned}
	\label{eq:ex_ez}
\end{equation}
with \( l_x \) and \( l_z \) respectively denoting the distance from the PML boundaries along the $x$ and $z$ directions:
\begin{equation}
	\begin{aligned}
		l_x &= \max(0, x_{bl} - x) + \max(0, x - x_{br}), \\
		l_z &= \max(0, z_{bt} - z) + \max(0, z - z_{bb}), 
	\end{aligned}
	\label{eq:lx_lz}
\end{equation}
where \( x_{bl}, x_{br}, z_{bt}, z_{bb} \) denote the coordinates of the left, right, top, and bottom boundaries, respectively. These profiles are zero within the physical domain and gradually increase inside the PML, ensuring a smooth transition and absorption of outgoing waves.

The decay strength \( c \) is frequency dependent and defined as
\begin{equation}
	c = a_0 \frac{\omega_0}{\omega L_{\text{PML}}^2},
	\label{eq:c}
\end{equation}
where \( \omega_0 \) is a reference angular frequency, \( \omega \) is the current frequency, \( L_{\text{PML}} \) is the PML thickness, and \( a_0 \) is a tunable scaling parameter. The inverse proportionality to \( \omega \) ensures that waves of different frequencies are damped appropriately over the same PML width.

Substituting Equations \eqref{eq:ex_ez}--\eqref{eq:c} into Equation \eqref{eq:scattered_helmholtz_PML} leads to an expanded form of Helmholtz equation in which the PML coefficients appear explicitly in each term. This shows how the spatially varying damping introduced by the PML influences both the differential operators and the source term of the scattered Helmholtz equation. The modified coefficients cause directional attenuation based on the distance to the domain boundaries, effectively simulating an unbounded medium while maintaining computational tractability.

We also derive an analytical approximation for the background wavefield in the presence of the PML (see \ref{app:PML}):
\begin{equation}
	u_0(\mathbf{x}) = \frac{i}{4} H_0^{(2)}\left(\frac{\omega}{v_0}  |\mathbf{x} - \mathbf{x}_f|\right) \exp\left[-\omega c \frac{(l_x^2+l_z^2)}{3 v_0}^{(3/2)}\right],
\end{equation}
where \( H_0^{(2)}(\cdot) \) is the zero-order Hankel function of the second kind, \( v_0 \) is the constant background velocity, and \( \mathbf{x}_f \) is the source location. This equation is exact within the domain of interest, and an approximation in the PML region.

\section{Methodology}
\label{sec:methodology}
	\subsection{Loss Function}
	
	The solution for $u_s(\mathbf{x})$ is obtained using a PINN framework. The loss function combines the PDE loss ($\mathcal{L}_{\text{PDE}}$) and a soft constraint term ($\mathcal{L}_{\text{C}}$):
	\begin{equation}
		\mathcal{L} = \mathcal{L}_{\text{PDE}} + \beta \mathcal{L}_{\text{C}}.
		\label{eq:loss_total}
	\end{equation}
	
	Without the PML, and when the velocity model is real-valued (non-attenuative), \( \mathcal{L}_{\text{PDE}} \) is the loss enforcing the real and imaginary parts of the solution to satisfy the scattered Helmholtz equation \citep{Alkhalifah2021}:
	\begin{equation}
		\begin{aligned}
			\mathcal{L}_{\text{PDE}} = \frac{1}{N} \sum_{j=1}^{N} \left( 
			\left| \nabla^2 u_s^r(\mathbf{x}_j) + \frac{\omega^2}{v^2(\mathbf{x}_j)} u_s^r(\mathbf{x}_j) + \omega^2 \delta m(\mathbf{x}_j) u_{0}^r(\mathbf{x}_j) \right|^2 
			\right. + \\
			\left. \left| \nabla^2 u_s^i(\mathbf{x}_j) + \frac{\omega^2}{v^2(\mathbf{x}_j)} u_s^i(\mathbf{x}_j) + \omega^2 \delta m(\mathbf{x}_j) u_{0}^i(\mathbf{x}_j) \right|^2 \right),
		\end{aligned}
		\label{eq:loss_pde}
	\end{equation}
	where $N$ is the number of collocation points. Note that the two terms in the above equation share the background wavefield $u_0$, which links the real and imaginary components of the scattered field.
	
Our numerical experiments with PINNs show that classical boundary reflections do not appear even without explicitly implementing an absorbing boundary condition. This aligns with observations reported by other researchers (e.g., \cite{Alkhalifah2021, Rasht-Behesht2022, wu2023}). We believe this is related to the spectral bias of PINNs and the fact that reflections generate solutions with higher oscillations. Nonetheless, in our applications, PMLs can be beneficial because they explicitly impose the coupling of the real and imaginary parts of the scattered wavefield, and enforce physical boundary conditions on the wavefield at the domain boundaries \citep{ wu2023}.
	
		Incorporating the PML in the scattered wavefield PDE (equation~\ref {eq:scattered_helmholtz_PML}), and defining the $\mathcal{L}_{\text{PDE}}$ as the mean squared difference between the two sides of that equation, the loss function takes the form (see \ref{app:PML}):  
\begin{equation}
	\mathcal{L}_{\text{PDE}} = \frac{1}{N} \sum \left( L_r(u_s^r, u_s^i)^2 + L_i(u_s^r, u_s^i)^2 \right),
	\label{eq:loss_pml}
\end{equation}
	where \( L_r \) and \( L_i \) represent the real and imaginary components of the loss, respectively, as functions of the real (\( u_s^r \)) and imaginary (\( u_s^i \)) parts of the scattered field. For implementation efficiency, we explicitly calculate the real and imaginary parts of the loss, as shown in  \ref{app:PML}, to avoid using complex numbers during training. With the addition of the PML, each term of the loss depends on both the real and imaginary components of the neural network output.

To prevent the trivial solution \(u_s = -u_0 \), the soft constraint term \( \mathcal{L}_{\text{C}} \) minimizes the solution close to the source where $ u_0 $ has a large magnitude \citep{huang2023b}. The term is calculated as:
	\begin{equation}
		\mathcal{L}_{\text{C}} = \frac{1}{N_{\text{C}}} \sum_{j=1}^{N_{\text{C}}} \left|u_s(\mathbf{x}_{\text{C},j}) \cdot \gamma(\mathbf{x}_{\text{C},j})\right|^2,
	\end{equation}
	where \( \mathbf{x}_{\text{C},j} \) are the \(N_{\text{C}}\) collocation points within the soft constraint  radius, and \( \gamma(\mathbf{x}_{\text{C},j}) \) is a scaling factor based on the distance from the source:
	\begin{equation}
		\gamma(\mathbf{x}_{\text{C},j}) =\sqrt{ \max\left(0, \frac{\lambda^2}{4} - r_f^2 \right)},
	\end{equation}
	where \(r_f\) represents the distance from the source, and \( \lambda \) is the wavelength.

	\subsection{Trainable Gabor Basis Functions}
	\label{sec:Gabor_functions}
	
When the output layer of a neural network is linear, the resulting wavefield is a linear combination of the outputs from the penultimate layer. By analogy, these outputs can be viewed as trainable basis functions that capture complex spatial patterns of the solution, and we will therefore employ the term \textit{trainable basis functions} in this work. Notice, however, that they do not strictly meet the mathematical definition of basis functions, as they can be linearly dependent.

	\begin{figure}[tbp]
	\centering
	\includegraphics[width=0.4\textwidth]{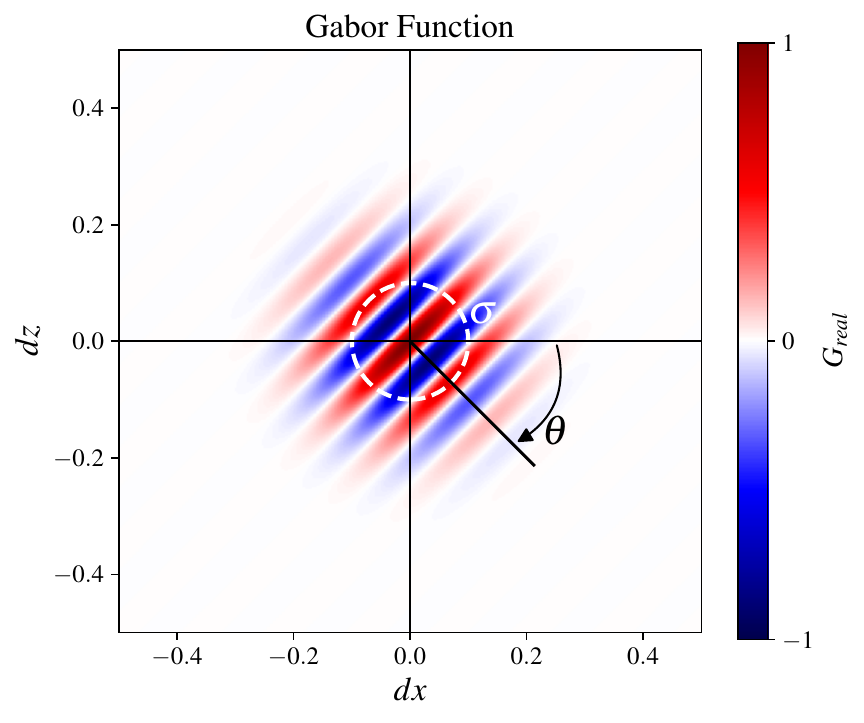}
\caption{Real part of a 10 Hz Gabor function displayed as a function of \( (d_x, d_z) \), where \( \sigma = 0.1 \) controls the Gaussian envelope width and \( \theta = -\pi/4 \) is the initial rotation angle of the wavefront.}
	\label{fig:Gabor_schematic}
\end{figure}
	
	\textbf{Gabor Function:} Gabor functions combine a Gaussian envelope with a sinusoidal carrier and present both local and oscillatory behaviors of the solution \citep{pinto2014}. Therefore, they are well-suited for wavefield simulations if used as explicit trainable basis functions.

	A conventional two-dimensional Gabor function centered at \( (\mu_x, \mu_z) \) is given by \citep{gabor1946}:
	\begin{equation}
		G(x, z) = A \exp \left( -\frac{x'^2 + z'^2}{2\sigma^2} \right) 
		\exp \left( i \left( \frac{\omega}{v} x' + \psi \right) \right),
	\end{equation}
	where \( (x', z') \) are rotated coordinates defined as:
	\begin{align}
		x' &= (x - \mu_x) \cos(\theta) + (z - \mu_z) \sin(\theta), \\
		z' &= -(x - \mu_x) \sin(\theta) + (z - \mu_z) \cos(\theta).
	\end{align}
	Here, $\omega$ is angular frequency, \( \sigma \) controls the spatial spread of the Gaussian envelope, \( \theta \) is the rotation angle of the Gabor function in the spatial domain, \( A \) is its peak magnitude, and \( \psi \) is the phase offset. The sinusoidal carrier represents the oscillatory behavior, while the Gaussian envelope ensures spatial localization.

	In \cite{Alkhalifah2024}, Gabor functions are used at the output of the penultimate layer in a multiplicative form, which means their main network learns \( A_p(\mathbf{x}) \), representing the magnitude of the \( p \)-th basis function at a given input location. To allow each basis function to contribute differently across spatial coordinates, the Gabor centers are also modeled as functions of the input coordinates, i.e., \( \mu(\mathbf{x}) \). They train an auxiliary network to output \( \mu(\mathbf{x}) \) for given collocation points.  
	
We define the local displacement vector as  
\begin{equation}
	\mathbf{d}(\mathbf{x}) = \mathbf{x} - \boldsymbol{\mu}(\mathbf{x}),
\end{equation}  
where \( \mathbf{d}(\mathbf{x}) = (d_x(\mathbf{x}), d_z(\mathbf{x})) \) denotes the horizontal and vertical distances from the Gabor function center \( \boldsymbol{\mu}(\mathbf{x}) \) in a local coordinate system (see Figure~\ref{fig:Gabor_schematic}). This displacement, when used with the Gaussian envelope, captures the spatial localization of the Gabor kernel and implicitly includes the effects of both the amplitude \( A_p(\mathbf{x}) \) and the center \( {\mu}_p(\mathbf{x}) \) for each mode \( p \).

Further simplifying the formulation by removing the phase offset, the real and imaginary components of our trainable Gabor basis functions are defined as:
\begin{align}
	G_{\text{real}}(d_x, d_z) &= \cos\left( \frac{\omega}{v} d_{x_\theta} \right) 
	\exp\left( -\frac{1}{2\sigma^2} \left( d_{x_\theta}^2 + d_{z_\theta}^2 \right) \right), \\
	G_{\text{imag}}(d_x, d_z) &= \sin\left( \frac{\omega}{v} d_{x_\theta} \right) 
	\exp\left( -\frac{1}{2\sigma^2} \left( d_{x_\theta}^2 + d_{z_\theta}^2 \right) \right),
\end{align}
where the transformed coordinates \((d_{x_\theta}, d_{z_\theta})\) are given by:
\begin{align}
	d_{x_\theta} &= d_x \cos(\theta) + d_z \sin(\theta), \\
	d_{z_\theta} &= -d_x \sin(\theta) + d_z \cos(\theta).
\end{align}

	\begin{figure}[tb]
		\centering
		\begin{minipage}{0.59\textwidth}
			\centering
			\includegraphics[width=\textwidth]{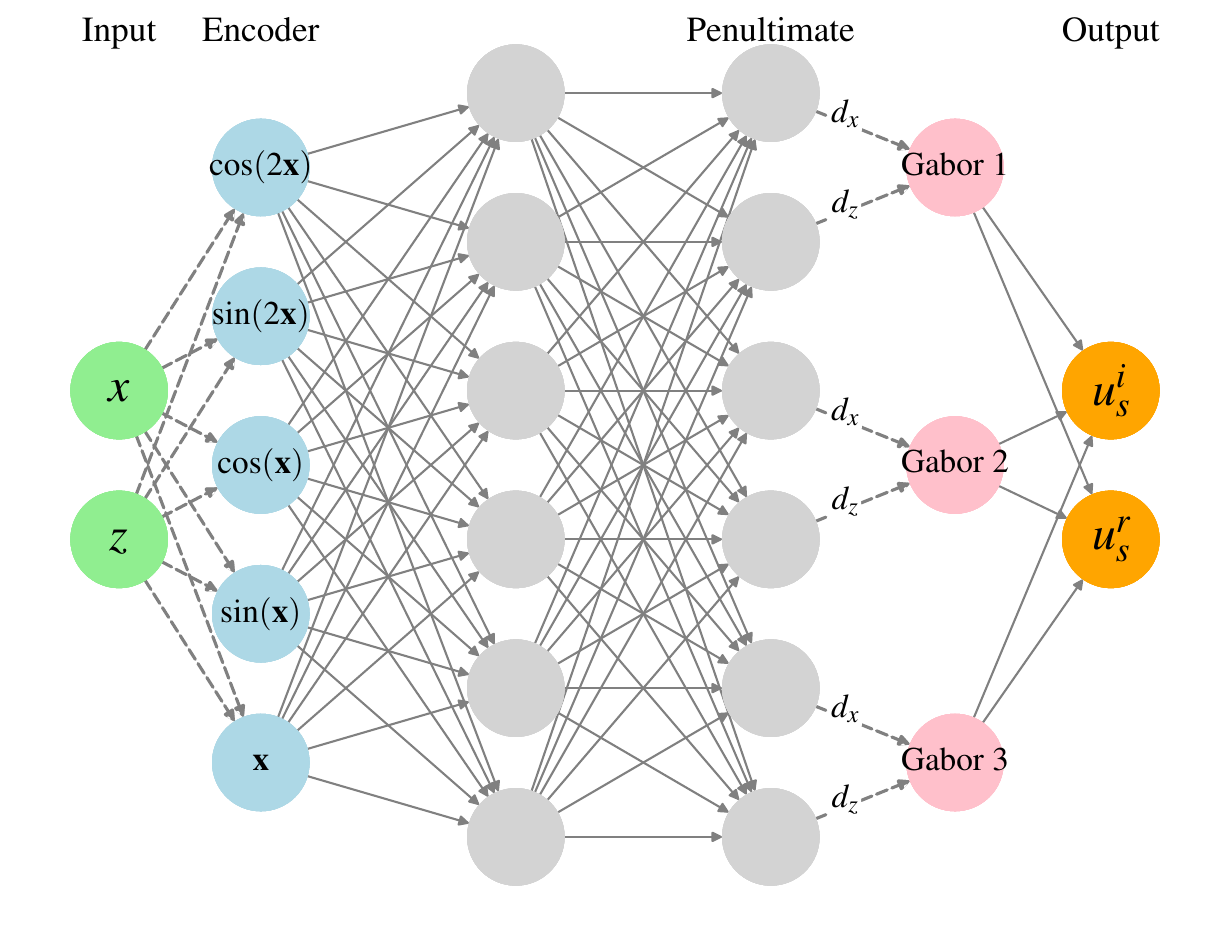}  % Part (b)
		\end{minipage}
		\caption{Schematic representation of the neural network architecture. Dashed connections have no associated weight. The network learns a mapping from the input \((x,z)\) to \((d_x,d_z)\) in the Gabor functions that produce the oscillatory behavior of the wave field \(u_s\).	}
		\label{fig:NN_architecture}
	\end{figure}
	
	\begin{figure}[htb]
		\centering
		\includegraphics[width=0.7\textwidth]{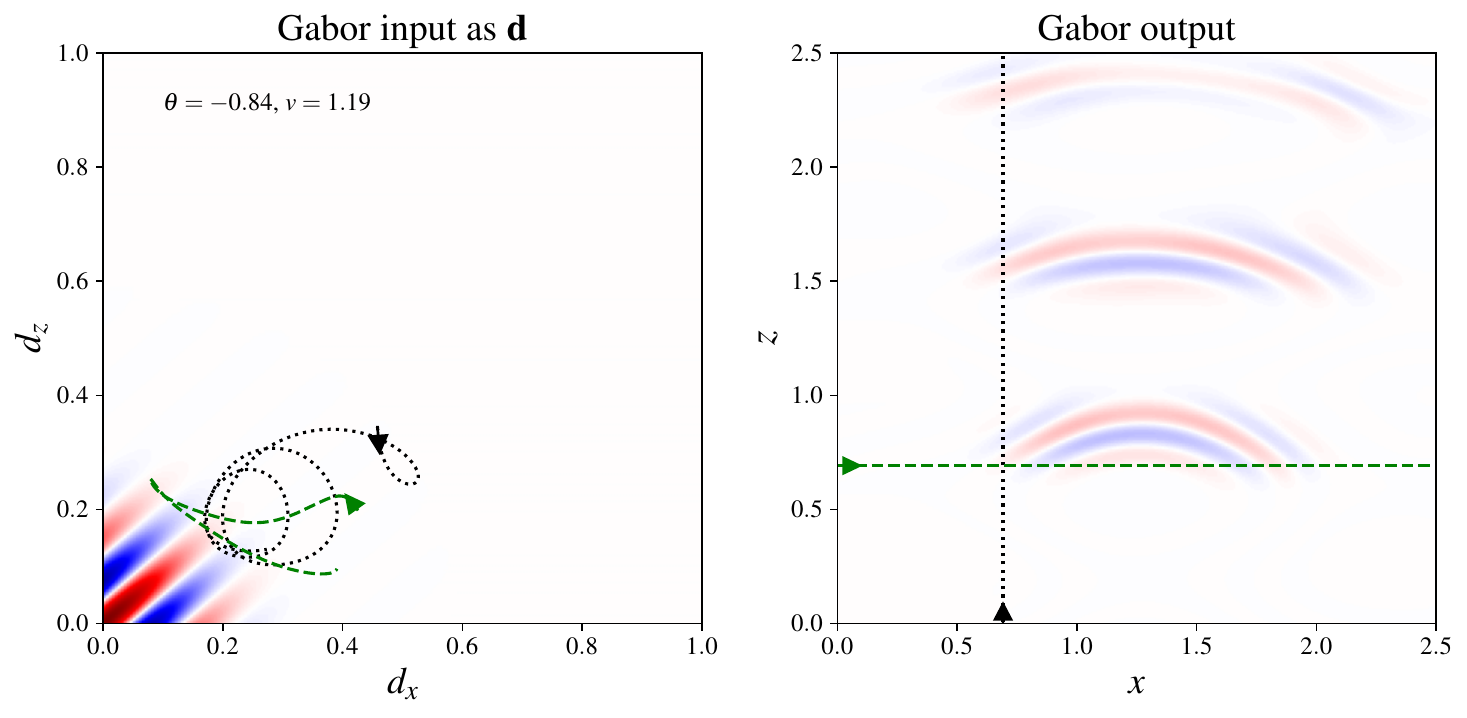}
		\caption{(a) Real part of the Gabor function plotted as a function of \( (d_x, d_z) \). The superimposed paths represent the trajectories produced by the network for vertical and horizontal sets of input collocation points \( (x, z) \). (b) Output values of this Gabor basis function, illustrating how the learned mapping from \( (x, z) \) to \( (d_x, d_z) \) enables the Gabor basis function to contribute to the wavefield solution.}
		\label{fig:gabor_mapping}
	\end{figure}

	\section{Implementation }
	\label{sec:implementation}
	A feed-forward neural network (NN) is trained to minimize the loss function.  The network's task is to learn the mapping from the input collocation points \( (x, z) \)  to the corresponding \( (d_x, d_z) \) values for each Gabor basis function. To implement it, pairs of neurons from the penultimate layer are used as \( d_x \) and \( d_z \), as shown in Figure \ref{fig:NN_architecture}. The other parameters of the Gabor functions are set as global parameters, meaning they do not vary with each input. This allows the network to learn only the transformation from the input coordinates \( (x, z) \) to the appropriate \( (d_x, d_z) \) coordinates corresponding to the required wavefield behavior. 
	
	The angular frequency \( \omega \) is fixed to its value in the objective Helmholtz equation, as it dictates the oscillatory behavior of the wavefield. We fix the standard deviation \( \sigma = 0.1 \), and constrain the range of \( d_x \) and \( d_z \) to the interval \( [0, 1] \), to ensure that the Gabor function decays to zero within the possible range of \( d_x \) and \( d_z \).
	
Since the relationship between the \(( d_x, d_z) \) coordinates and the original spatial coordinates \( (x, z) \) is nonlinear, there is no need for the parameters \( \theta \) (the rotation angle) or \( \frac{\omega}{v} \) (the wavelength) to directly match those in the target wavefield solution. Instead, these parameters are treated as trainable variables, initialized to the same values in all neurons. Specifically, we initialize \( v \) to \( v_0 \) and \( \theta \) to \( -\pi/4 \), which orients the Gabor functions diagonally across the domain. Given the constrained domain \( (d_x, d_z) \in [0, 1] \), this diagonal initialization places the Gabor wavefronts along the panel's center and ensures that the directional response is not biased toward axis-aligned features (Figure~\ref{fig:gabor_mapping}a). 

We tested alternative initializations by assigning random values to \( \theta \) within the valid range, but found that such randomness led to degraded convergence and accuracy in the numerical experiments. This suggests that a consistent initialization, aligned diagonally, provides a more effective and stable starting point for training.

Although our method introduces Gabor functions, it does not increase the total number of trainable parameters compared to a standard PINN with the same architecture. Each Gabor function contributes only two trainable parameters, and their inclusion halves the number of weights in the output layer because the number of Gabor functions is half the number of neurons in the penultimate layer. This balance preserves the overall trainable parameter count.

	When for a given input location \( (x, z) \), the Gabor function \( G \) is nonzero for the calculated \(( d_x, d_z) \), that basis function contributes to the solution at that point in the wavefield. Since the Gabor functions already incorporate the oscillatory behavior of the wavefield, the mapping from \( (x, z) \) to \( (d_x, d_z )\) learned by the network can be smooth and easily learned, allowing for efficient training and accurate representation of the wavefield. Figure \ref{fig:gabor_mapping} illustrates the mapping between the physical coordinates \( (x, z) \) and the corresponding Gabor function coordinates \( (d_x, d_z) \), highlighting how the neural network utilizes the Gabor basis function to represent the wavefield.
	
	In Figure \ref{fig:gabor_mapping}a, the real part of a Gabor function is plotted as a function of \( (d_x, d_z) \), with all other parameters fixed. The surface represents the value of the Gabor function across different points in the \( (d_x, d_z) \) space. Superimposed are two paths, which represent the \( (d_x, d_z) \) trajectories that the neural network produced for two sets of input collocation points; one set corresponding to vertical points and the other to horizontal points in the \( (x, z) \) domain. Figure \ref{fig:gabor_mapping}b shows the corresponding values produced by the Gabor basis function for these two sets of input collocation points in the physical space. The network learns to map each input \( (x, z) \) to a specific trajectory in the \( (d_x, d_z) \) space, allowing the Gabor basis function to contribute appropriately to the wavefield solution for each location point.
	
	The activation function used in hidden layers of the NN is chosen as,
	\begin{equation}
		\phi(\mathbf{h}) = \sin(\mathbf{h}),
	\end{equation}
	which is well-suited for modeling oscillatory wavefields \citep{huang2021,wong2022learning,song2022versatile,buzaev2024hybrid}.
	
	To enhance the representation of the input spatial coordinates \( (x, z) \), we use positional encoding to map the input into a higher-dimensional space, enabling the network to better capture oscillatory behaviors \citep{Vaswani2017, huang2021}. We use a simple sinusoidal encoding function:
	\begin{equation}
		\mathbf{E}(\mathbf{x}) = \left[x,z, \sin(2^k x), \cos(2^k x), \sin(2^k z), \cos(2^k z) \right]_{k=0}^{K},
		\label{eq:encoder}
	\end{equation}
	where \( \mathbf{E}(\mathbf{x}) \in \mathbb{R}^{2(K+1)} \) and \( K \) is the maximum frequency used for the encoding. A schematic representation of the encoder layer is shown in Figure \ref{fig:NN_architecture}. 
	
	\begin{figure}[]
		\centering
		\includegraphics[width=\textwidth]{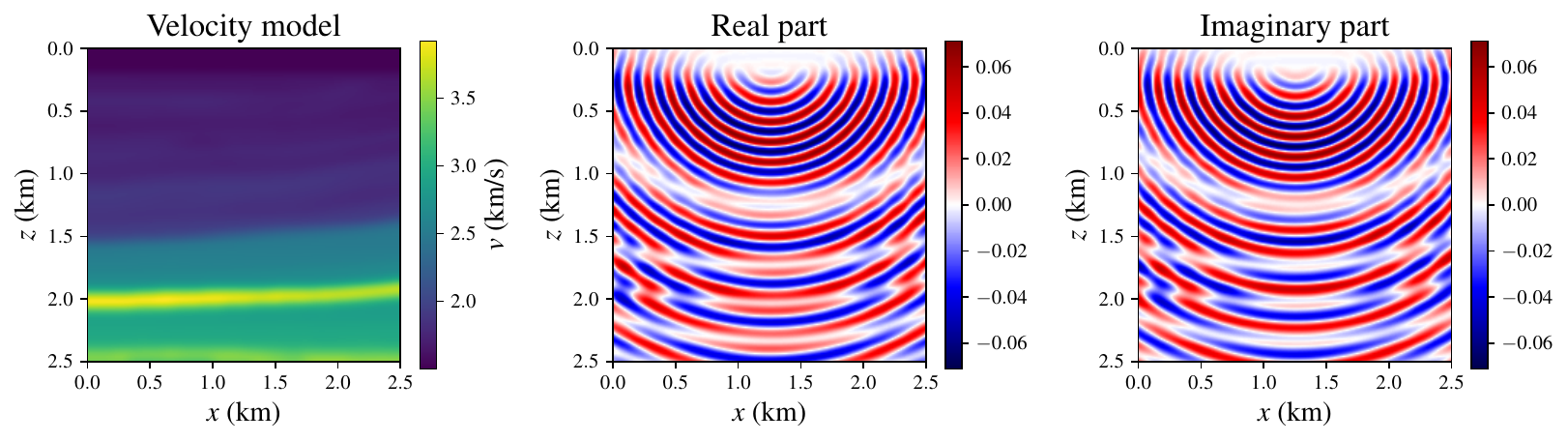}  
		\caption{A simple velocity model and the finite difference (FD) modeled 10Hz wave field using a fine grid spacing. We use this simulation result to calculate the error of different PINN predictions in Test 1. }
		\label{fig:10Hz_fd_wavefoeld}
	\end{figure}

\begin{figure}[tp]
\centering
\begin{minipage}{0.32\textwidth}
	\centering
	\includegraphics[width=\textwidth]{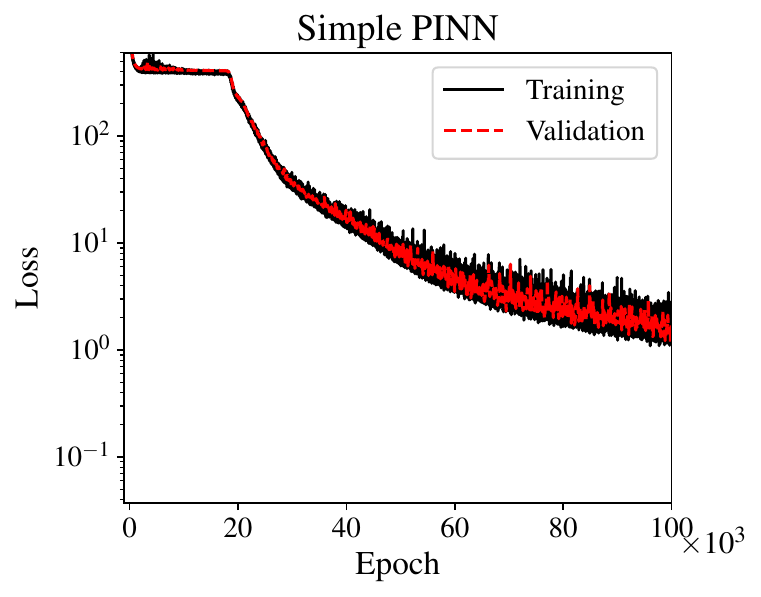}
\end{minipage}
\begin{minipage}{0.32\textwidth}
	\centering
	\includegraphics[width=\textwidth]{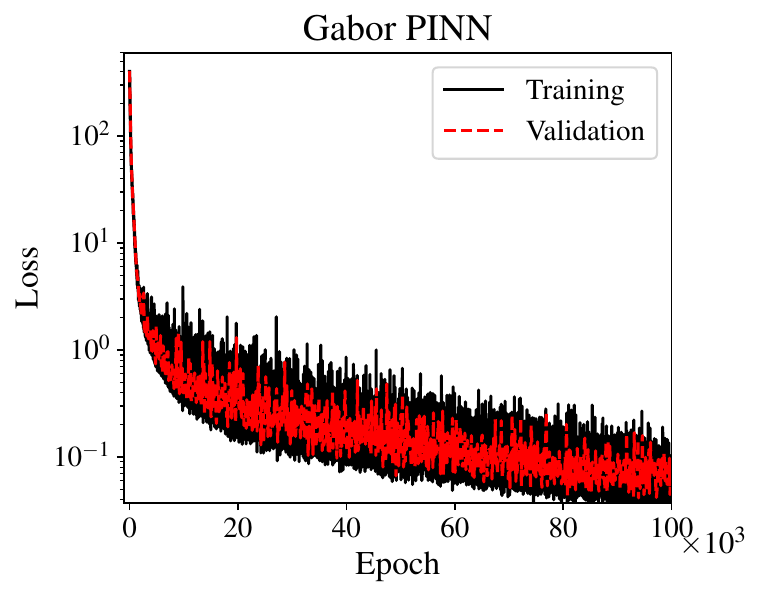}
\end{minipage}
\begin{minipage}{0.32\textwidth}
	\centering
	\includegraphics[width=\textwidth]{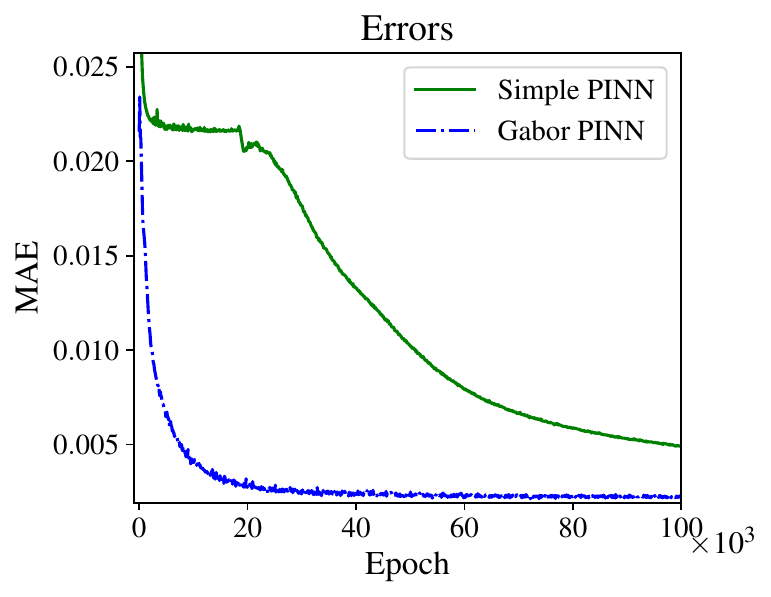}
\end{minipage}

\caption{
	Convergence comparison between a simple PINN and the proposed Gabor-PINN for simulating a 10 Hz wavefield (Test 1) in the velocity model from Figure~\ref{fig:10Hz_fd_wavefoeld}. The only difference between the models is the proposed addition of Gabor functions. (a) Loss evolution for the simple PINN, (b) loss evolution for the Gabor-PINN, and (c) validation error for both methods. The Gabor-PINN shows faster convergence and improved accuracy.
}

\label{fig:loss_10Hz}
\end{figure}
	
\begin{figure}[htbp]
		\centering
		\begin{minipage}{0.30\textwidth}
			\centering
			\includegraphics[width=\textwidth]{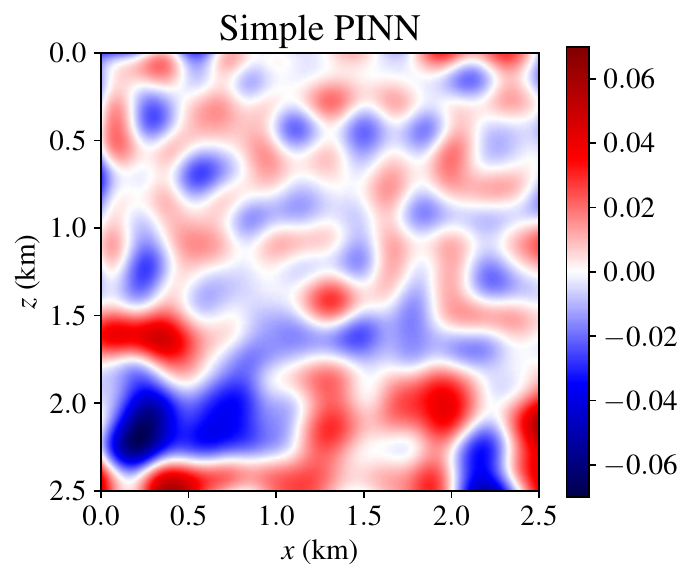}
		\end{minipage}
		\begin{minipage}{0.30\textwidth}
			\centering
			\includegraphics[width=\textwidth]{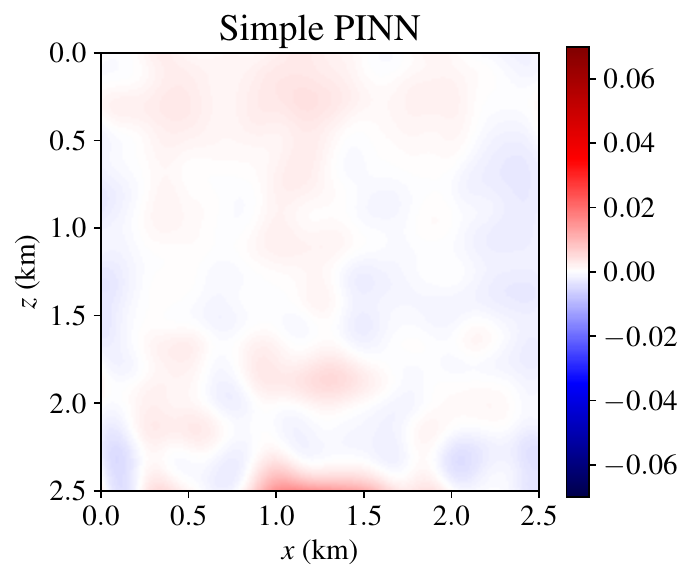}
		\end{minipage}
		\begin{minipage}{0.30\textwidth}
			\centering
			\includegraphics[width=\textwidth]{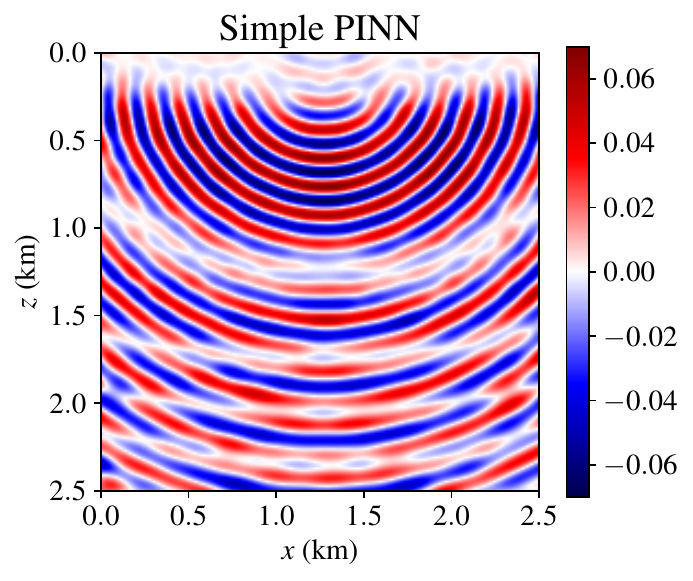}
		\end{minipage}
		\\
		\begin{minipage}{0.30\textwidth}
			\centering
			\includegraphics[width=\textwidth]{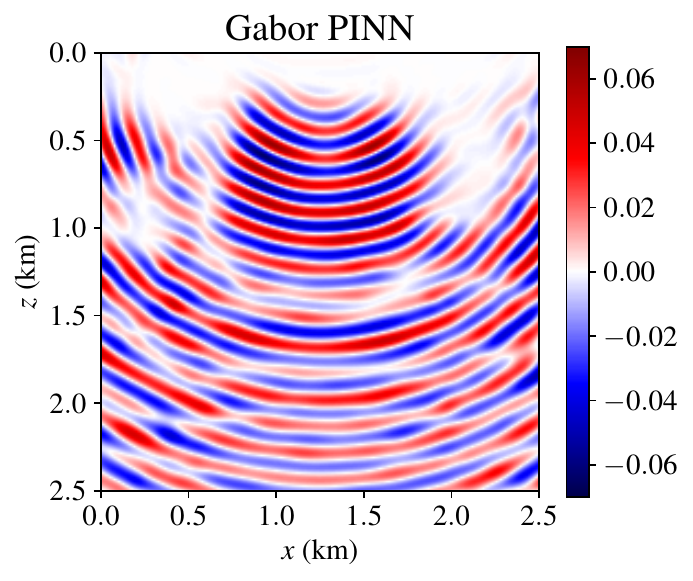}
			\caption*{(a) 500 epochs}
		\end{minipage}
		\begin{minipage}{0.30\textwidth}
			\centering
			\includegraphics[width=\textwidth]{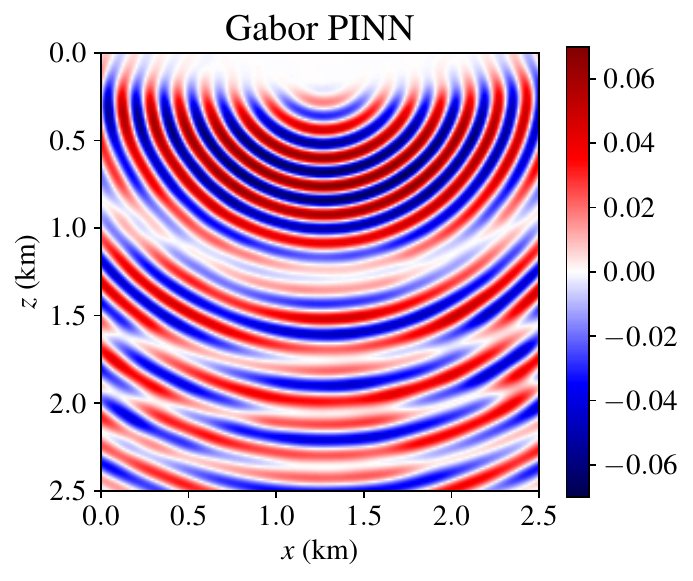}
			\caption*{(b) 10,000 epochs}
		\end{minipage}
		\begin{minipage}{0.30\textwidth}
			\centering
			\includegraphics[width=\textwidth]{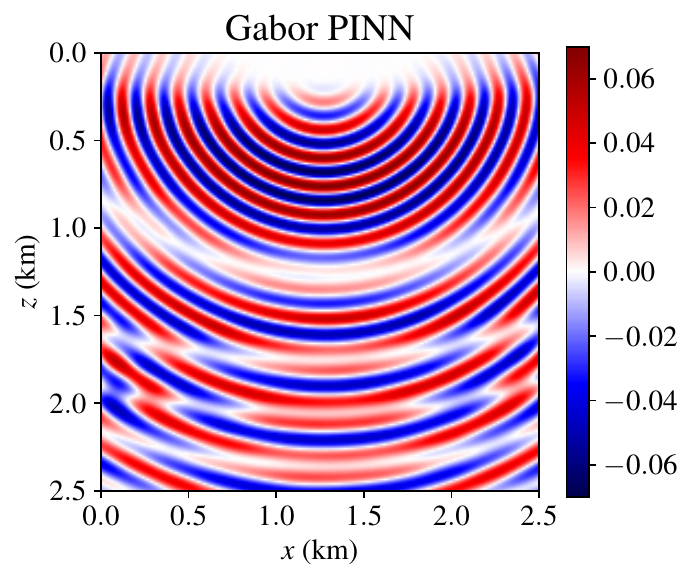}
			\caption*{(c) 100,000 epochs}
		\end{minipage}
\caption{Real part of the predicted 10 Hz wavefield from the proposed simple PINN and the proposed Gabor-PINN at different training epochs (500, 10,000, and 100,000) for Test 1. Both methods share the same training settings, number of trainable parameters, and network architecture, except for the inclusion of Gabor functions in the Gabor-PINN. The Gabor-PINN exhibits faster convergence. The reference finite difference (FD) simulated wavefield is shown in Figure~\ref{fig:10Hz_fd_wavefoeld}.}
\label{fig:predictions_10Hz}

	\end{figure}

\section{Numerical Results}
\label{sec:numerical_results}
We evaluate the performance and properties of the proposed Gabor-PINN method through six numerical experiments. These tests are organized into three categories: (1) ablation studies to assess the effect of incorporating Gabor functions, (2) studies on the inclusion of the perfectly matched layer (PML) in the training process, and (3) comparisons with other methods that incorporate Gabor functions into PINNs. 

The experiments are conducted using three distinct velocity models and cover four source frequencies ranging from 4 Hz to 20 Hz. This diverse set of scenarios allows us to analyze the method's convergence behavior, modeling accuracy, robustness to architecture and initialization, and computational efficiency.

\subsection{Ablation Studies}

An ablation study systematically removes or modifies components of a model to evaluate their individual contributions to overall performance. To assess the specific role of the proposed Gabor functions, we conduct a series of ablation studies. In each test, we compare a \textit{Gabor-PINN} model that incorporates our implementation of the Gabor functions with an otherwise identical baseline \textit{simple-PINN} model that excludes them. This comparison isolates the effect of the Gabor functions on model accuracy, convergence, and robustness.  Table~\ref{tab:hyperparam_summary} summarizes the network configurations, number of trainable parameters, Multiply-Accumulate operations (MACs), and the number of collocation points per epoch used in each test. MACs measure the total number of elementary operations required for one forward pass through the network and serve as an indicator of computational complexity.

\subsubsection{Test 1: Convergence at 10 Hz}

We begin with a simple velocity model shown in Figure~\ref{fig:10Hz_fd_wavefoeld}, and generate a reference 10~Hz scattered wavefield using finite difference (FD) modeling. The FD solution is computed on a grid with four times finer spacing in each direction and includes a large perfectly matched layer (PML) to minimize boundary reflections. The FD wavefield is downsampled to a \(200 \times 200\) grid for comparison with PINN predictions.

\textbf{Simple PINN: }We first train a standard PINN consisting of three hidden layers, each with 64 neurons, and employ an encoder layer with \( K = 3 \). During training, we sample 2,601 collocation points per epoch. The loss is defined as in Equations~\eqref{eq:loss_total} and \eqref{eq:loss_pde}, and the network is trained for 100{,}000 epochs using an exponentially decaying learning rate (initially 0.001, decaying to 0.0003). The learning rate decay helps improve stability, and has been shown to enhance training in PINNs (e.g., \cite{bihlo2024improving, xu2024preprocessing}). As shown in Figure~\ref{fig:loss_10Hz}a, the simple PINN begins converging only after 20{,}000 epochs.

\textbf{Gabor-PINN: } We add 32 Gabor functions to the network, as described in the implementation section, without changing the number of trainable parameters. All other settings, including the architecture, training parameters, collocation points, and learning rate, remain identical to the simple PINN. Figure~\ref{fig:loss_10Hz}b shows that the Gabor-PINN achieves much faster convergence and lower loss throughout training.

Figure~\ref{fig:loss_10Hz}c shows the validation errors, evaluated against the FD reference wavefield. Notably, the Gabor-PINN reaches the lowest error of the simple PINN in just 7{,}000 epochs. Figure~\ref{fig:predictions_10Hz} further illustrates the qualitative improvement, comparing wavefield predictions at epochs 500, 10{,}000, and 100{,}000. The Gabor-PINN consistently outperforms the simple PINN at each stage of training.

\subsubsection{Test 2: Higher-frequency Wavefield (20 Hz)}

To evaluate performance in a more challenging scenario, we simulate a 20~Hz scattered wavefield using the same velocity model. The higher frequency induces stronger oscillations and a more complex wavefield structure. Here, we use a larger network with three hidden layers of 128 neurons each, and set the positional encoding parameter to \( K = 5 \). Each training step uses 90{,}601 collocation points. Detailed hyperparameters and MAC counts for both models are provided in Table~\ref{tab:hyperparam_summary}.

\begin{figure}[tb]
		\centering
		\begin{subfigure}[b]{0.32\textwidth}
			\includegraphics[width=\textwidth]{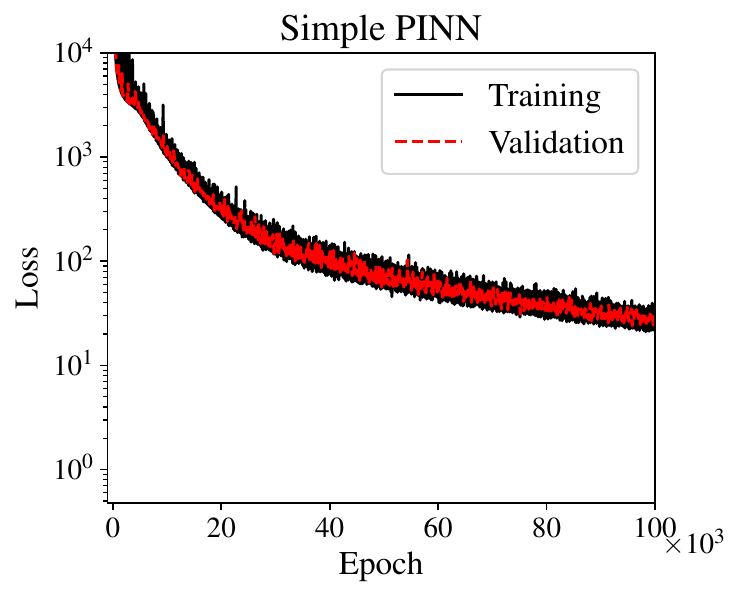}
			%			\caption{\cite{Alkhalifah2024} method}
		\end{subfigure}
		\begin{subfigure}[b]{0.32\textwidth}
			\includegraphics[width=\textwidth]{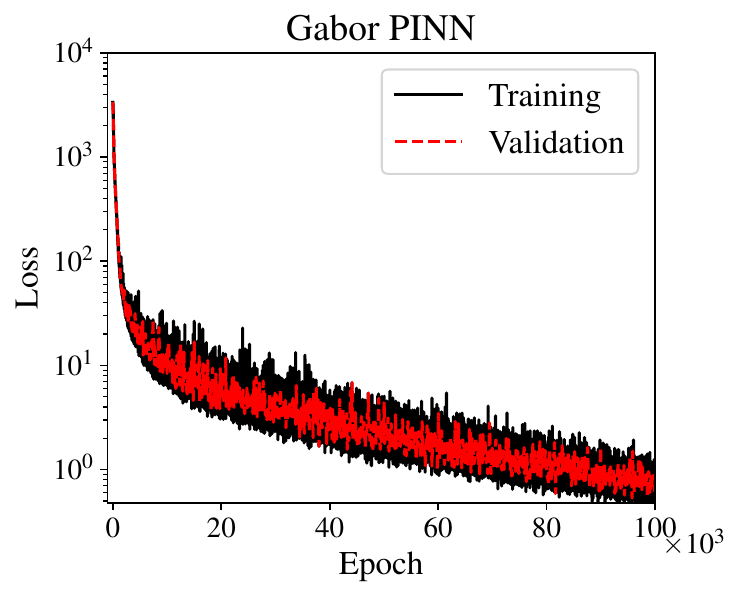}
			%			\caption{Loss: Proposed Gabor-PINN}
		\end{subfigure}
		\begin{subfigure}[b]{0.32\textwidth}
			\includegraphics[width=\textwidth]{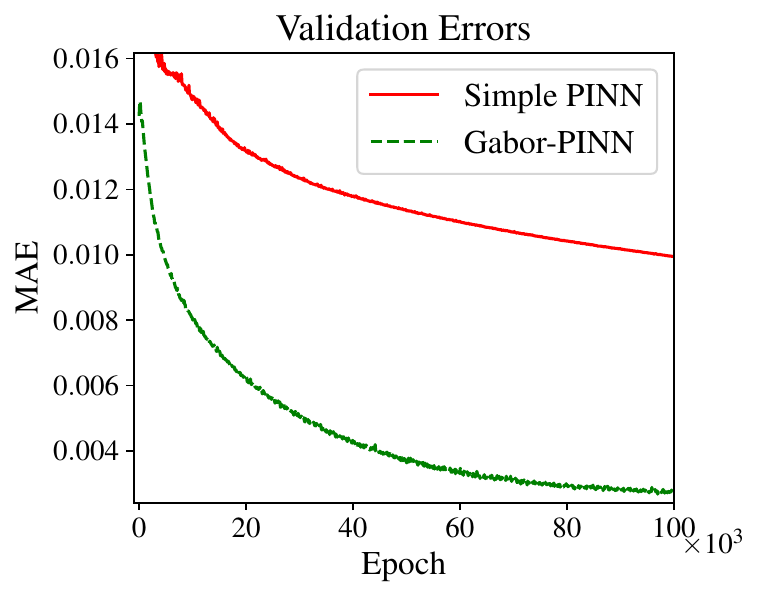}
			%			\caption{Validation errors}
		\end{subfigure}
		
		\begin{subfigure}[b]{0.32\textwidth}
			\includegraphics[width=\textwidth]{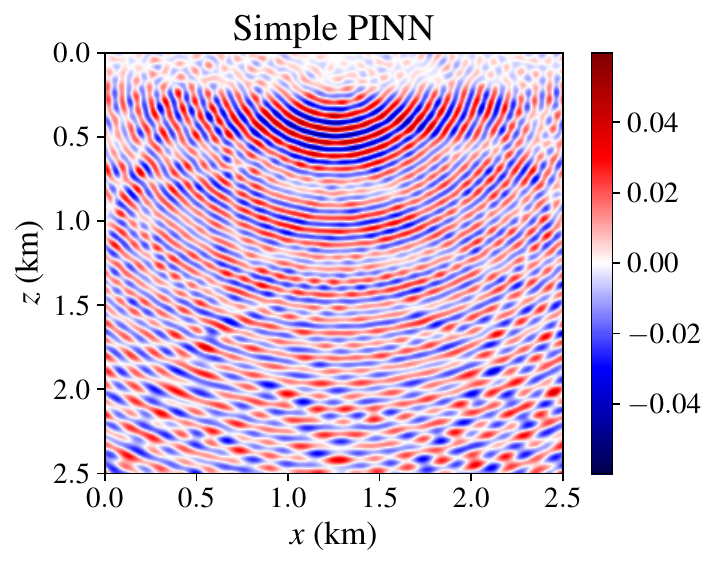}
			%			\caption{Prediction: \cite{Alkhalifah2024} method}
		\end{subfigure}
		\begin{subfigure}[b]{0.32\textwidth}
			\includegraphics[width=\textwidth]{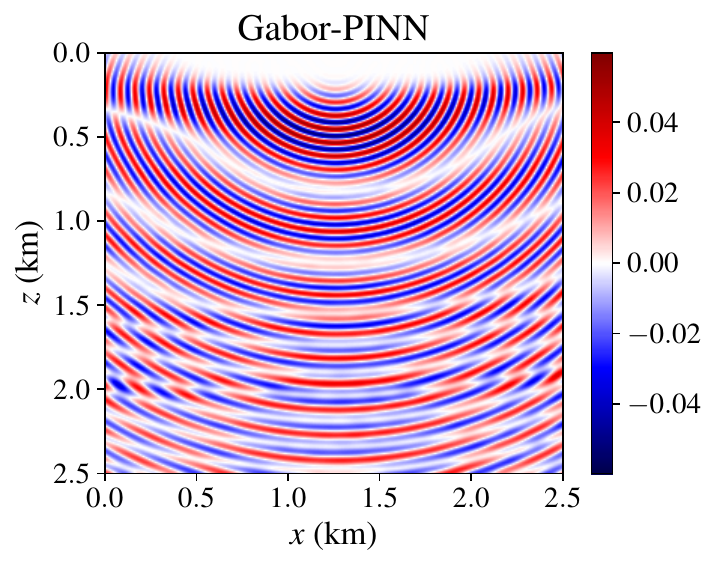}
			%			\caption{Prediction: Proposed Gabor-PINN}
		\end{subfigure}
		\begin{subfigure}[b]{0.32\textwidth}
			\includegraphics[width=\textwidth]{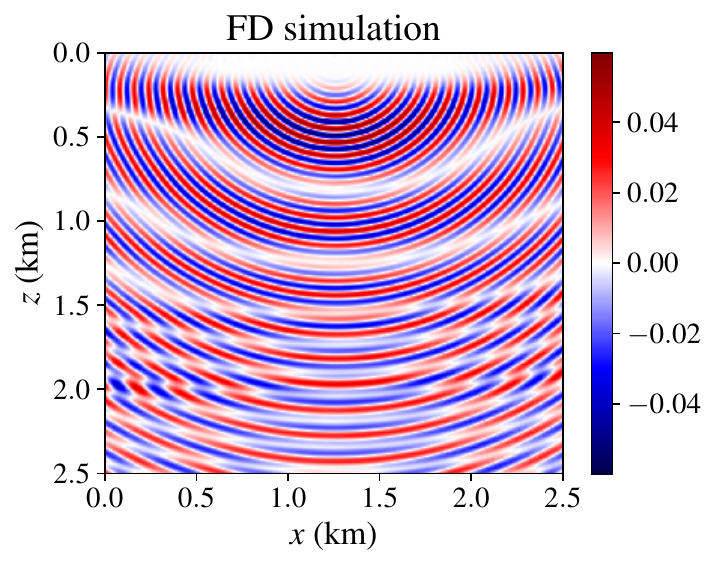}
			%			\caption{Reference wavefield (FD)}
		\end{subfigure}
		
\caption{
	Comparison of a simple PINN and the proposed Gabor-PINN for simulating a 20 Hz scattered wavefield (Test 2) in the velocity model from Figure~\ref{fig:10Hz_fd_wavefoeld}. The validation error curves demonstrate significantly lower errors for the Gabor-PINN, highlighting its improved accuracy in high-frequency wavefield modeling.
}

\label{fig:20Hz_test}
	\end{figure}

\textbf{Simple PINN: }Figure~\ref{fig:20Hz_test}a shows the loss curve for the simple PINN, trained using the same procedure as in Test 1. The model shows slow convergence and higher final loss in this high-frequency problem.

\textbf{Gabor-PINN: } We add 64 Gabor functions to the network while keeping all other settings identical. As seen in Figure~\ref{fig:20Hz_test}b, the Gabor-PINN converges faster and to significantly lower training and validation losses. The validation error in Figure~\ref{fig:20Hz_test}c confirms the improved accuracy of the Gabor-PINN.

Figures~\ref{fig:20Hz_test}d-e show the real parts of the predicted wavefields and the reference FD solution on a \(300 \times 300\) grid. The Gabor-PINN reproduces the wavefield more accuratly than the simple PINN.

\subsection{Analysis of PML Inclusion in Complex Velocity Models}

Here, we examine the impact of incorporating a perfectly matched layer (PML) in the training process of PINNs. By comparing the performance with and without the PML, we assess its contribution to convergence, stability, and accuracy.

\subsubsection{Test 3: Marmousi Velocity Model}

\begin{figure}[tph]
		\centering
		\begin{subfigure}[b]{0.32\textwidth}
			\includegraphics[width=\textwidth]{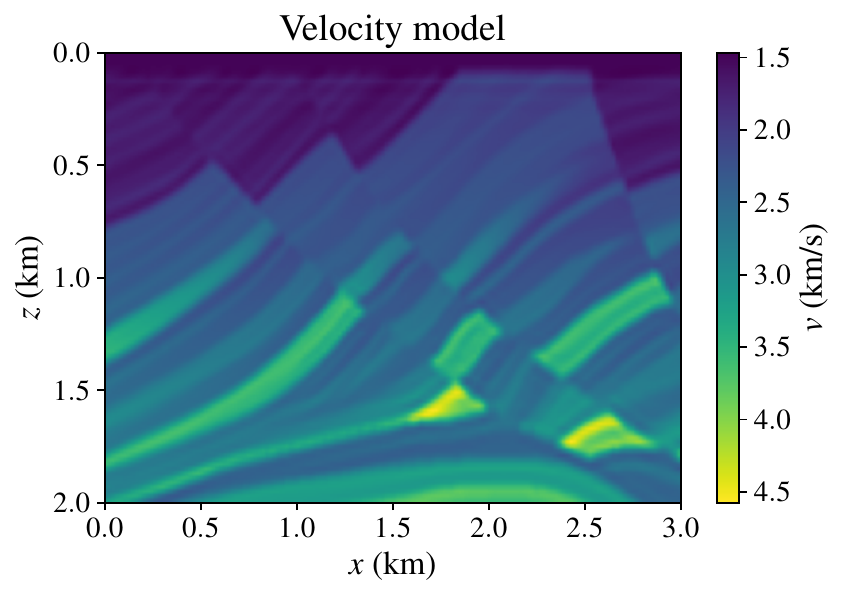}
			%			\caption{\cite{Alkhalifah2024} method}
		\end{subfigure}
		\begin{subfigure}[b]{0.32\textwidth}
			\includegraphics[width=\textwidth]{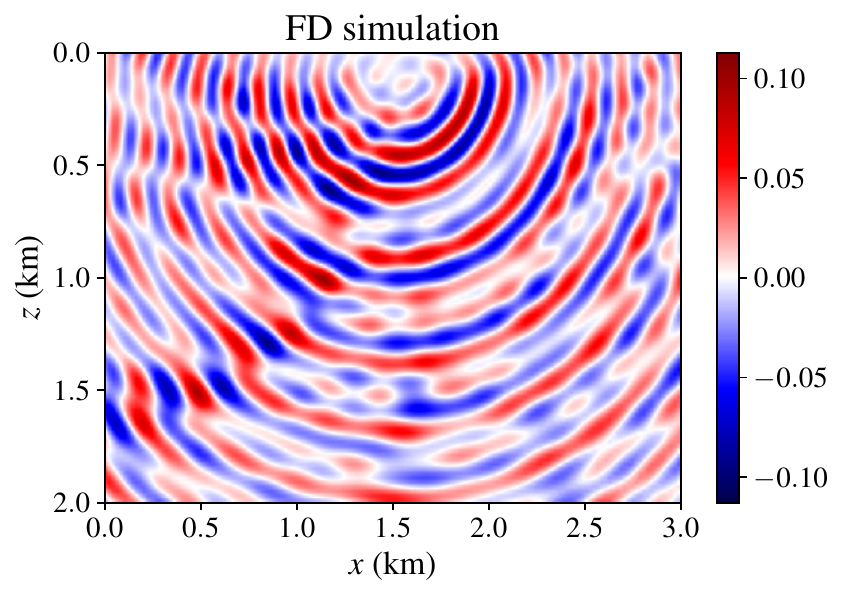}
			%			\caption{Loss: Proposed Gabor-PINN}
		\end{subfigure}
		\\
		\begin{subfigure}[b]{0.32\textwidth}
			\includegraphics[width=\textwidth]{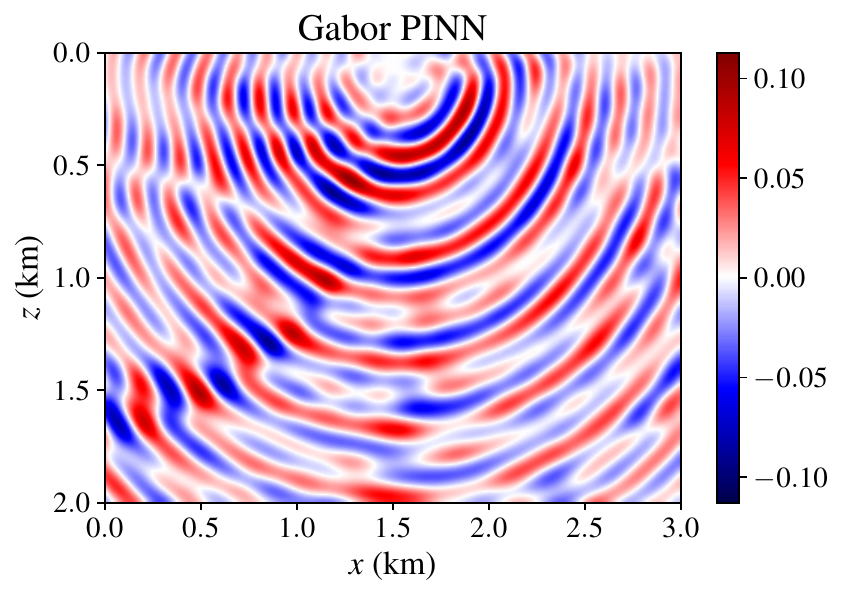}
			%			\caption{Validation errors}
		\end{subfigure}
		\begin{subfigure}[b]{0.32\textwidth}
			\includegraphics[width=\textwidth]{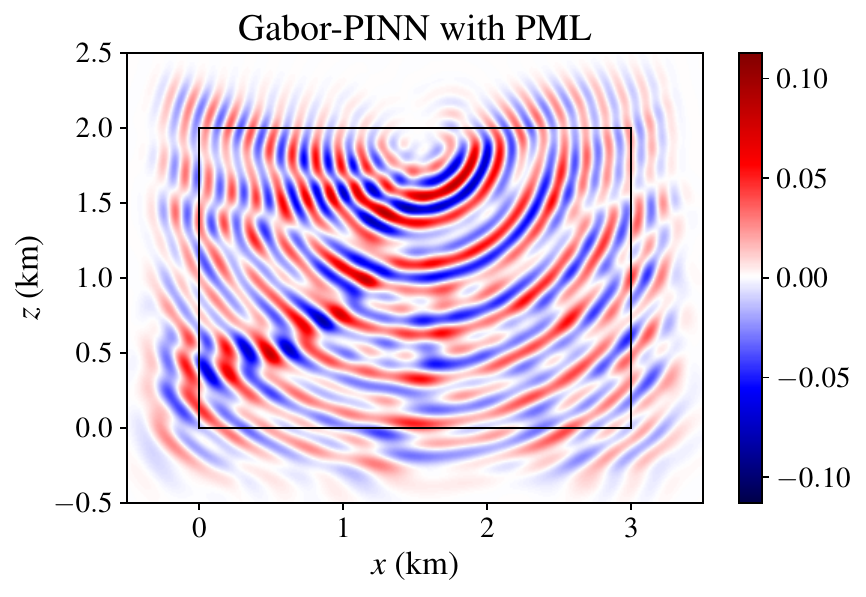}
			%			\caption{Prediction: \cite{Alkhalifah2024} method}
		\end{subfigure}
		
\caption{
	The Marmousi velocity model (Test 3) is shown in panel (a), and the reference finite difference (FD) wavefield in panel (b). Panels (c) and (d) display the Gabor-PINN wavefield predictions without and with PML, respectively. The corresponding loss and validation error curves are presented in the next figure.
}

\label{fig:Marmousi_prediction}
	\end{figure}

In this test, we evaluate the effect of the PML implementation using a portion of the benchmark Marmousi velocity model for a 10 Hz wavefield modeling. The relatively nonsmooth Marmousi model introduces high-contrast velocity variations, providing a challenging scenario for wavefield modeling. Figure~\ref{fig:Marmousi_prediction}a shows the velocity model, while Figure~\ref{fig:Marmousi_prediction}b presents the reference finite difference (FD) wavefield, presented on a 100 by 150 grid (using four times finer grid spacing for computation). The neural network architecture consists of 4 hidden layers, each with 128 neurons, and an encoder layer with \( K = 3 \). The network is trained for 100,000 epochs with an exponentially decaying learning rate, starting at 0.002 and decaying to 0.0007. Detailed hyperparameters and MAC counts for both models are provided in Table~\ref{tab:hyperparam_summary}.

\begin{figure}[h]
		\centering
		\begin{subfigure}[b]{0.32\textwidth}
			\includegraphics[width=\textwidth]{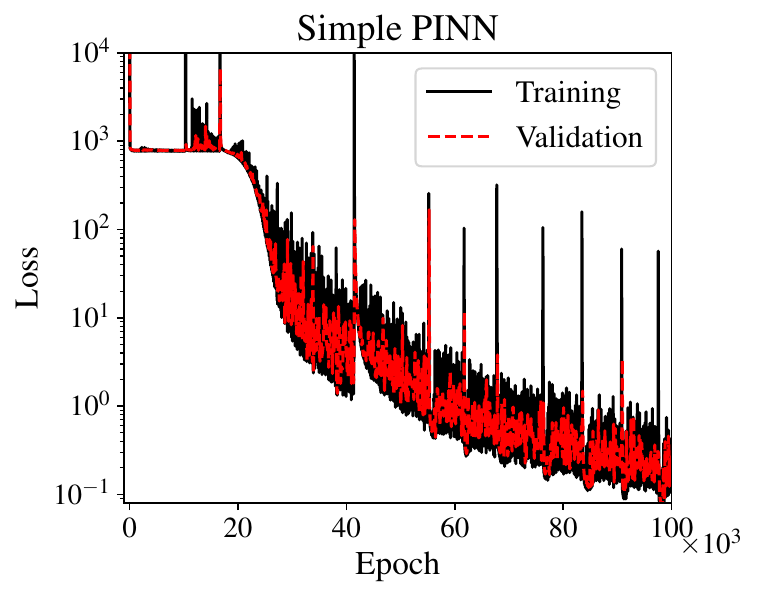}
			%			\caption{\cite{Alkhalifah2024} method}
		\end{subfigure}
		\begin{subfigure}[b]{0.32\textwidth}
			\includegraphics[width=\textwidth]{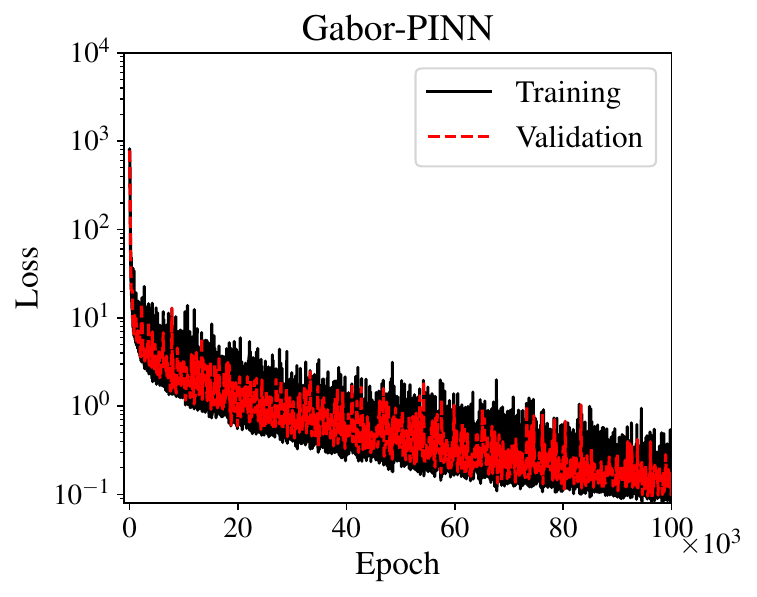}
			%			\caption{Loss: Proposed Gabor-PINN}
		\end{subfigure}
		\\
		\begin{subfigure}[b]{0.32\textwidth}
			\includegraphics[width=\textwidth]{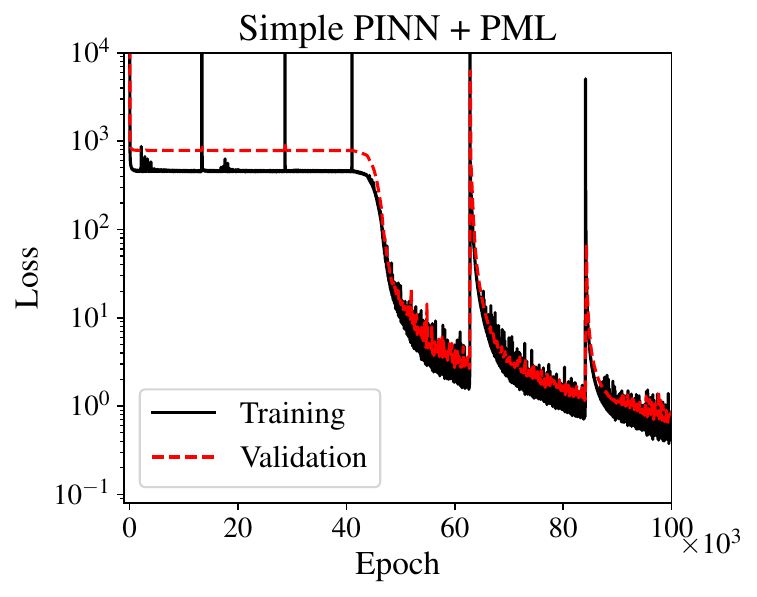}
			%			\caption{Validation errors}
		\end{subfigure}
		\begin{subfigure}[b]{0.32\textwidth}
			\includegraphics[width=\textwidth]{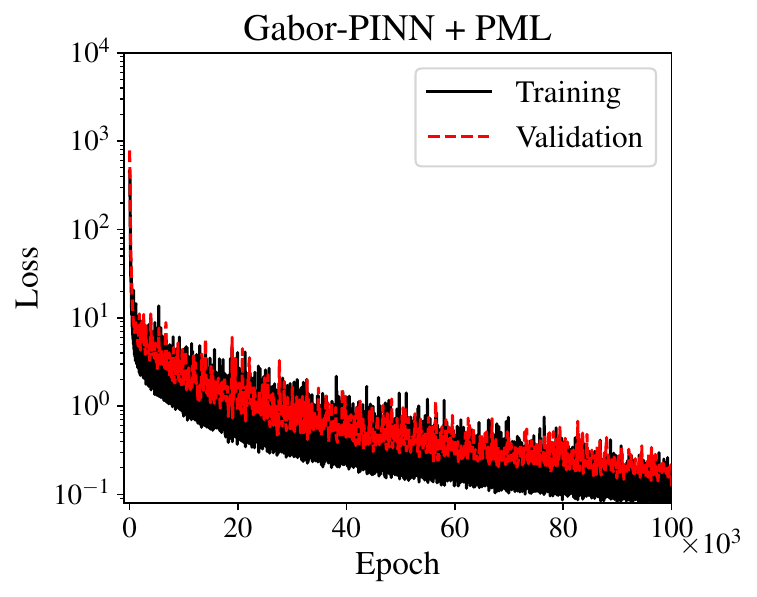}
			%			\caption{Prediction: \cite{Alkhalifah2024} method}
		\end{subfigure}
%		\begin{subfigure}[b]{0.32\textwidth}
%			\includegraphics[width=\textwidth]{20Hz_prediction_Gabor.pdf}
%			%			\caption{Prediction: Proposed Gabor-PINN}
%		\end{subfigure}
\\
		\begin{subfigure}[b]{0.55\textwidth}
			\includegraphics[width=\textwidth]{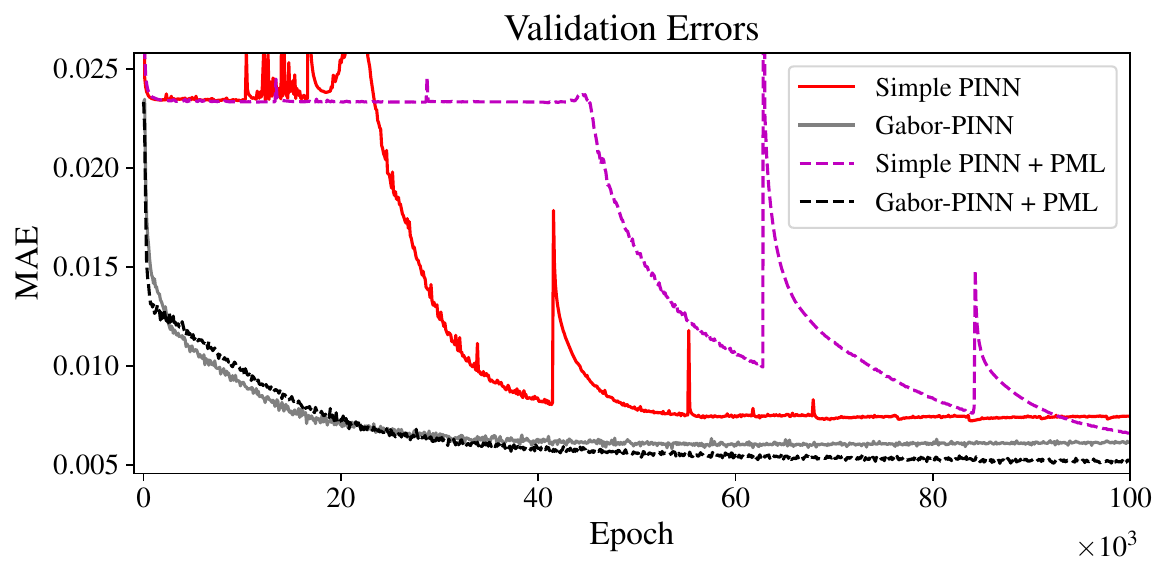}
			%			\caption{Reference wavefield (FD)}
		\end{subfigure}
		
\caption{Comparison of training and validation losses for the simple PINN and proposed Gabor-PINN methods with and without PML (Test 3, related to Figure~\ref{fig:Marmousi_prediction}). The Gabor-PINN demonstrates better convergence and accuracy, while the simple PINN shows a significant delay in convergence. Validation errors for both methods are shown, illustrating the improved final accuracy when PML is applied.}
\label{fig:Marmousi_losses}
	\end{figure}

We train both the simple PINN and our Gabor-PINN with and without the PML implementation. First, the networks are trained using the loss function based on equations \eqref{eq:loss_total} and \eqref{eq:loss_pde}, which do not include the PML. During training, we use 15,251 randomly varying collocation points in each step. Figure~\ref{fig:Marmousi_losses} shows the evolution of training and validation losses, as well as the validation errors compared to the reference FD wavefield. The simple PINN starts to converge after approximately 22,000 epochs, while the Gabor-PINN converges almost immediately and achieves higher accuracy.

Next, we define the loss using equations \eqref{eq:loss_total} and \eqref{eq:loss_pml}, incorporating the proposed definition for \( u_0 \) to include the PML. We add a PML with a thickness of $ L_{\text{PML}} = 0.5 \text{ km}$, using \( a_0 = 1 \). To account for the larger domain due to the PML region, we increase the number of collocation points proportionally, using 30,351 collocation points in each step. Our implementation does not increase the training time, except for the additional time required due to the larger number of collocation points. With the inclusion of PML, the convergence of the simple PINN is delayed until approximately 44,000 epochs (Figure~\ref{fig:Marmousi_losses}), whereas the Gabor-PINN's convergence is hardly affected. Both methods, however, achieve higher final accuracies with the PML included, as the validation errors in Figure~\ref{fig:Marmousi_losses} indicate.

Figure~\ref{fig:Marmousi_prediction} also shows the Gabor-PINN predictions with and without the PML. The wavefield decays successfully within the PML region, confirming the effectiveness of the PML implementation.

\subsubsection{Test 4: Overthrust Velocity Model}  

\begin{figure}[htbp]
		\centering
		\begin{minipage}{0.48\textwidth}
			\centering
			\includegraphics[width=\textwidth]{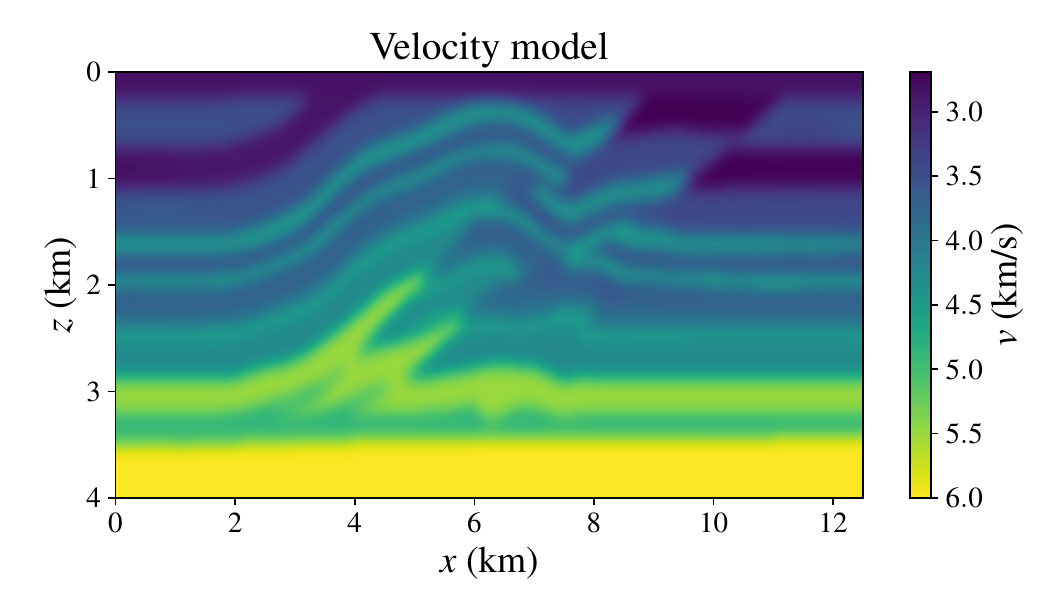}
			\caption*{(a)}
		\end{minipage}
		\begin{minipage}{0.25\textwidth}
			\centering
			\includegraphics[width=\textwidth]{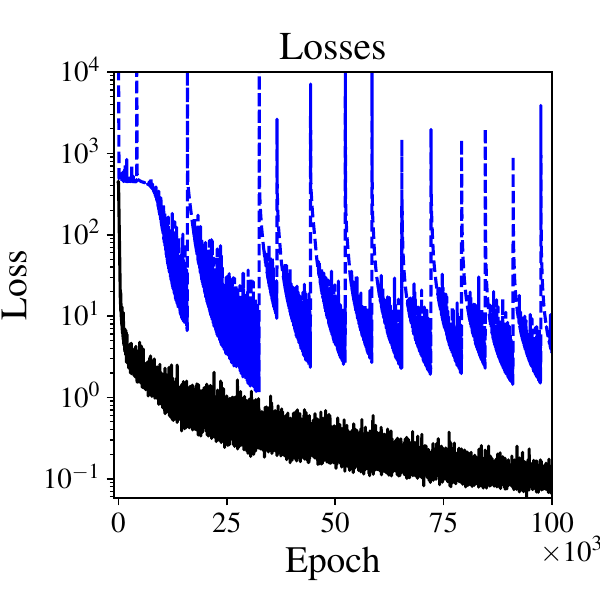}
			\caption*{(b)}
		\end{minipage}
				\begin{minipage}{0.25\textwidth}
			\centering
			\includegraphics[width=\textwidth]{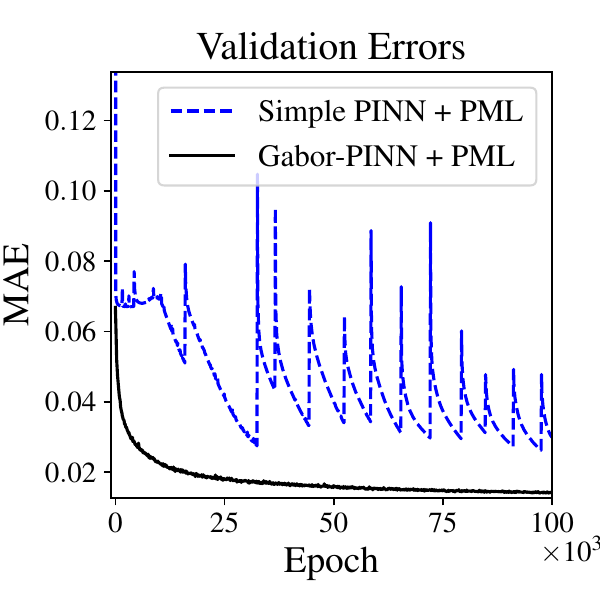}
			\caption*{(c)}
		\end{minipage}
		\\
		\begin{minipage}{0.48\textwidth}
			\centering
			\includegraphics[width=\textwidth]{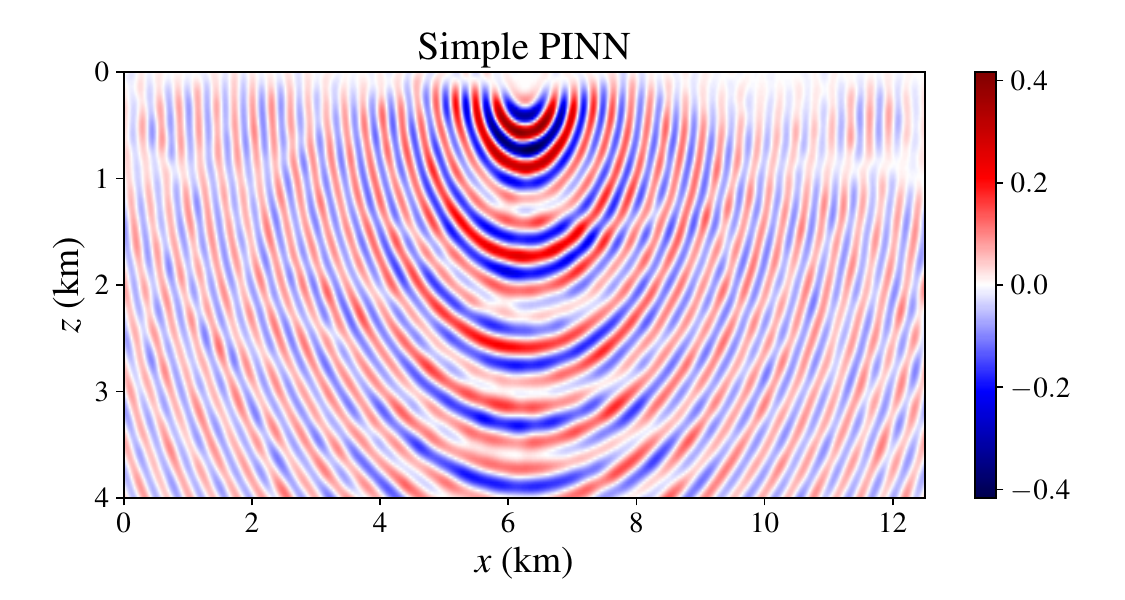}
			\caption*{(d)}
		\end{minipage}
		\begin{minipage}{0.48\textwidth}
			\centering
			\includegraphics[width=\textwidth]{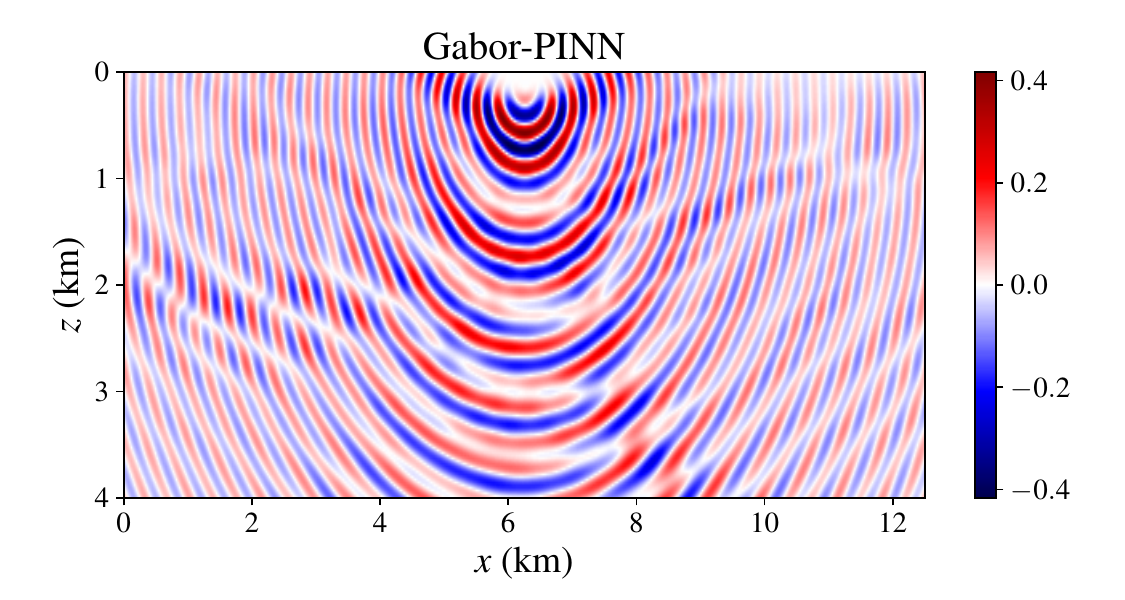}
			\caption*{(e)}
		\end{minipage}
				\\
		\begin{minipage}{0.49\textwidth}
			\centering
			\includegraphics[width=\textwidth]{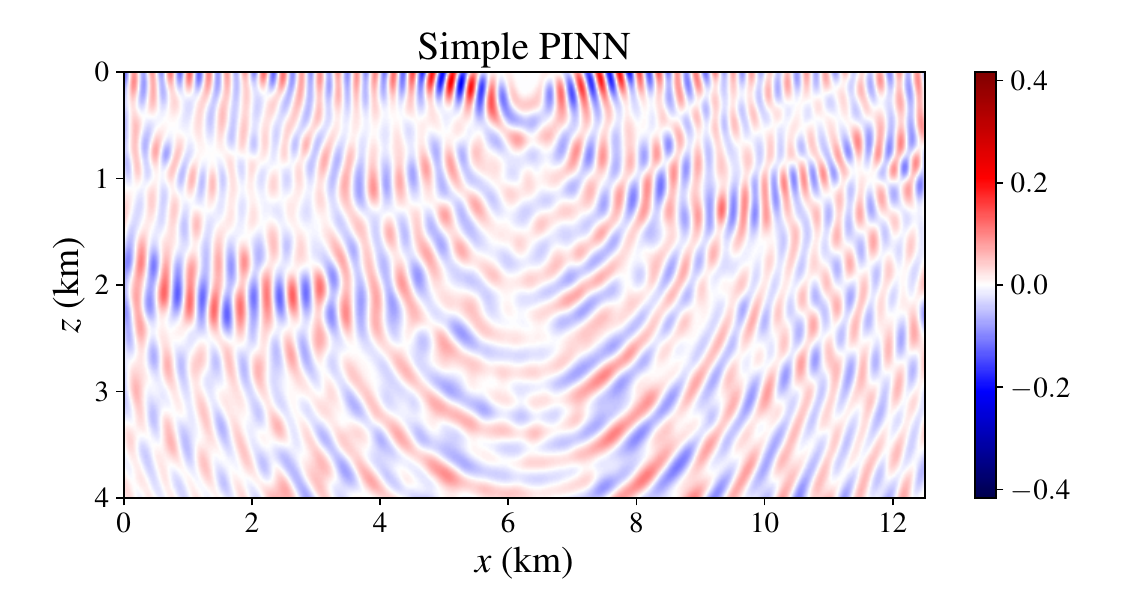}
			\caption*{(f)}
		\end{minipage}
		\begin{minipage}{0.49\textwidth}
			\centering
			\includegraphics[width=\textwidth]{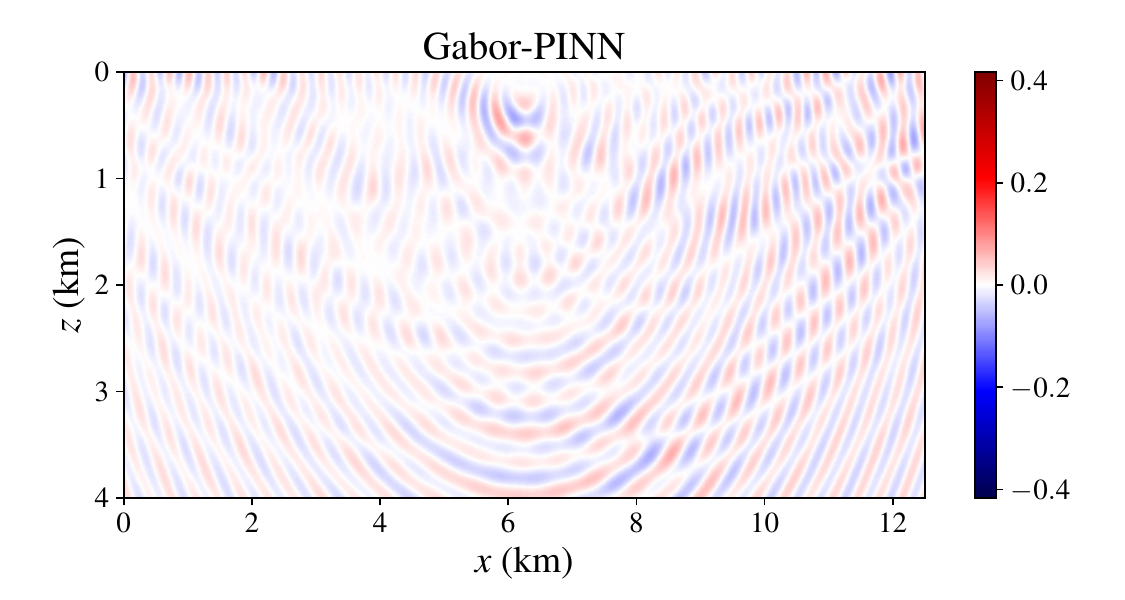}
			\caption*{(g)}
		\end{minipage}
		\caption{Test 4, (a) the Overthrust velocity model, (b) training loss and (c) validation error evolution for the simple PINN and proposed Gabor-PINN. The simple PINN exhibits unstable training, while the Gabor-PINN shows stable convergence. (d) and (e) show wavefield predictions on validation grid points for the simple PINN and Gabor-PINN, respectively. (f) and (g) present the prediction differences with the finite difference reference wavefield. The Gabor-PINN prediction is more accurate but still affected by low-frequency bias.}
		\label{fig:test3_OVE_results}
	\end{figure}

In this test, we evaluate the performance of our Gabor-PINN on the complex Overthrust velocity model. This model, depicted in Figure~\ref{fig:test3_OVE_results}a, has similarities to the subsurface in the Canadian foothills, featuring two overthrust faults. It is larger and more challenging than the previous models.  

We design a neural network with 4 hidden layers, each containing 128 neurons (resulting in 64 Gabor functions for our Gabor-PINN) and positional encoding with \( K = 3 \). The loss function is defined based on equations \eqref{eq:loss_total} and \eqref{eq:loss_pml}, which include the PML. We use a PML with a thickness of $ L_{\text{PML}} = 0.5 \text{ km}$, using \( a_0 = 1 \). The network is trained using the same learning rate schedule as in the previous test.  

Figures~\ref{fig:test3_OVE_results}b and c show the evolution of training loss and validation error for the simple PINN and the proposed Gabor-PINN. The simple PINN becomes unstable during training, resulting in frequent jumps in the loss and validation error. In contrast, the proposed Gabor-PINN demonstrates stable convergence.  

Figures~\ref{fig:test3_OVE_results}d and e show the wavefield predictions from the simple PINN and Gabor-PINN on the validation grid points. Panels (f) and (g) present the difference between the predictions and the finite difference (FD) reference wavefield. The proposed Gabor-PINN prediction is significantly more accurate than the simple PINN. However, the prediction still misses several fine details due to the low-frequency bias of the learned function. The sharpness of the solution may be improved by implementing a hybrid optimization scheme to reduce the spectral bias, which will be the focus of our future studies.

\begin{table}[htbp]
	\centering
	\small
	\caption{Summary of network hyperparameters, computational complexity represented by Multiply-Accumulate operations (MACs), positional encoding parameter \( K \), and number of collocation points per epoch (full batch). Parameters and MACs are in thousands (k). Note that adding our Gabor functions does not change the number of trainable parameters.}
	\renewcommand{\arraystretch}{1.1}
	\resizebox{\textwidth}{!}{%
		\begin{tabular}{lcccccc} 
			\toprule
			\textbf{Test} & \textbf{Layers} & \textbf{Neurons} & \textbf{\(K\)} & \textbf{Params (k)} & \textbf{MACs (k)} & \textbf{Collocation Points} \\ 
			\midrule
			Test 1 (10 Hz) & 3 & 64 & 3 & 9.66 & 9.57 (no Gabor) & 2,601 \\
			& & & & & 9.76 (with Gabor) & \\
			\midrule
			Test 2 (20 Hz) & 3 & 128 & 5 & 36.74 & 36.54 (no Gabor) & 90,601 \\
			& & & & & 36.93 (with Gabor) & \\
			\midrule
			Test 3 (Marmousi) & 4 & 128 & 3 & 52.22 & 51.97 (no Gabor) & 15,251 (no PML) \\
			& & & & & 52.35 (with Gabor) & 30,351 (with PML) \\
			\midrule
			Test 4 (Overthrust) & 4 & 128 & 3 & 52.22 & 51.97 (no Gabor) & 108,741 (with PML) \\
			& & & & & 52.35 (with Gabor) & \\
			\bottomrule
		\end{tabular}
	}
	\label{tab:hyperparam_summary}
\end{table}

\subsection{Comparative Study with Existing Gabor-PINN Methods}
\label{sec:comparison}

In this section, we compare our proposed method with two existing Gabor-based PINN formulations, corresponding to the work of \cite{Alkhalifah2024} and \cite{huang2023b}. For fairness and consistency, we adopt the hyperparameters that each method originally optimized for these specific wavefield simulation problems.

\subsubsection{Test 5: Sensitivity to Weight Initializations}
This test evaluates the robustness of our proposed Gabor-PINN compared to the implementation by \cite{Alkhalifah2024}, focusing on the effect of random initializations of the neural network weights. Note that the weights of our Gabor functions are initialized as explained in the Implementation section. The experiment is conducted for a 4~Hz wavefield using the simple velocity model shown in Figure~\ref{fig:10Hz_fd_wavefoeld}.

For this test, \cite{Alkhalifah2024} uses positional encoding with $K=3$, five layers with 32 neurons in the first four and 16 neurons in the last layer, followed by 16 Gabor functions, and accompanied by an auxiliary network of 8 neurons. Our Gabor-PINN uses a smaller network with three hidden layers of 32 neurons each and the same positional encoding and number of Gabor  functions as \cite{Alkhalifah2024}.

\begin{figure}[tb]
	\centering
	\begin{subfigure}[b]{0.32\textwidth}
		\includegraphics[width=\textwidth]{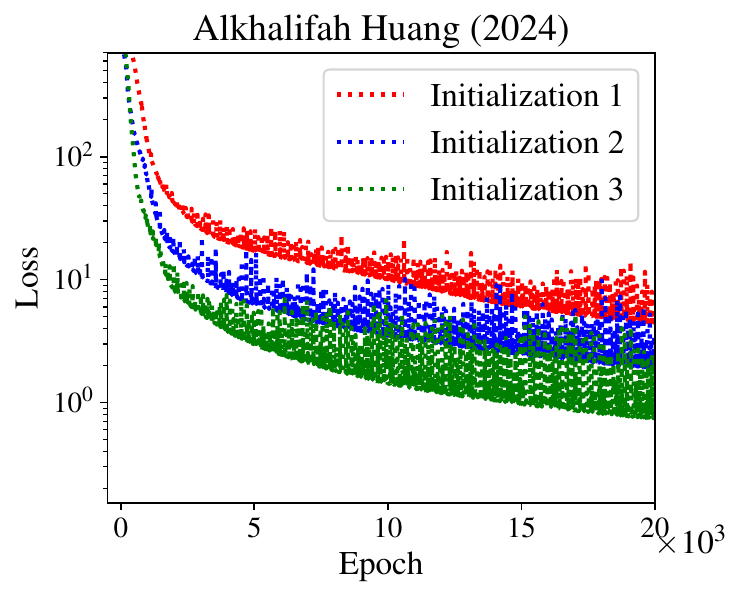}
		%			\caption{\cite{Alkhalifah2024} method}
	\end{subfigure}
	\begin{subfigure}[b]{0.32\textwidth}
		\includegraphics[width=\textwidth]{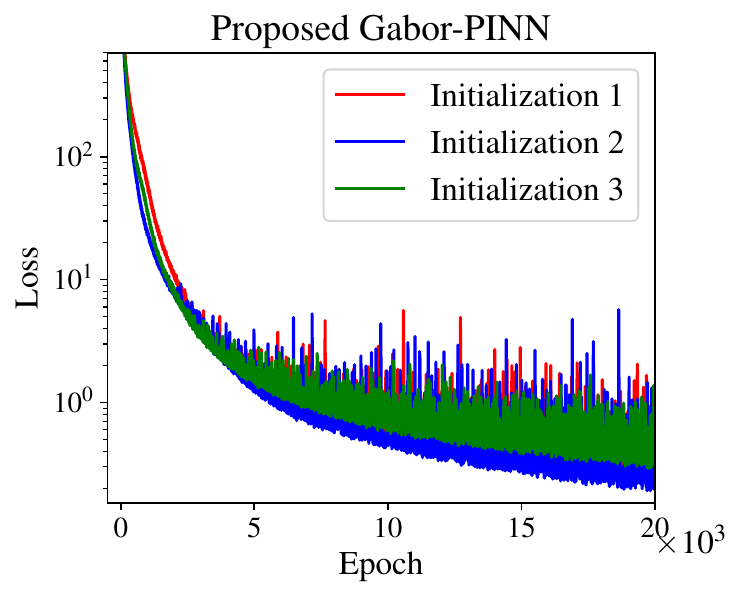}
		%			\caption{Loss: Proposed Gabor-PINN}
	\end{subfigure}
	\begin{subfigure}[b]{0.32\textwidth}
		\includegraphics[width=\textwidth]{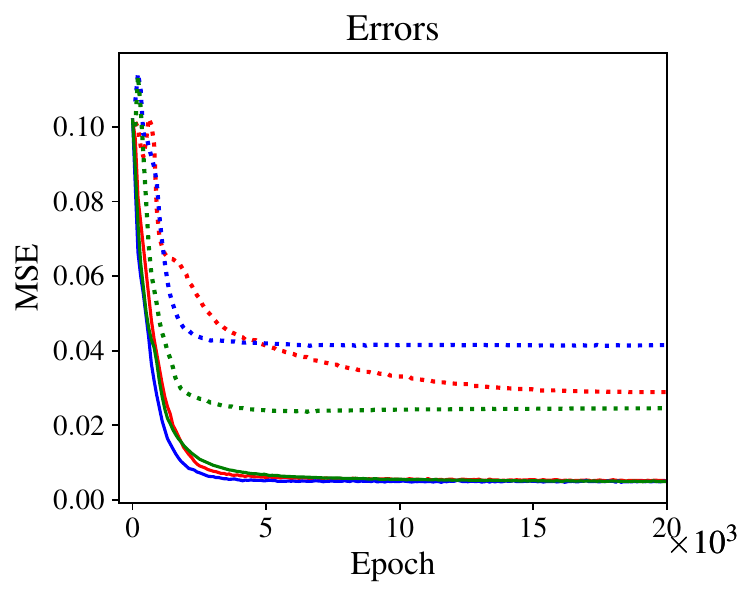}
		%			\caption{Validation errors}
	\end{subfigure}
	
	\begin{subfigure}[b]{0.32\textwidth}
		\includegraphics[width=\textwidth]{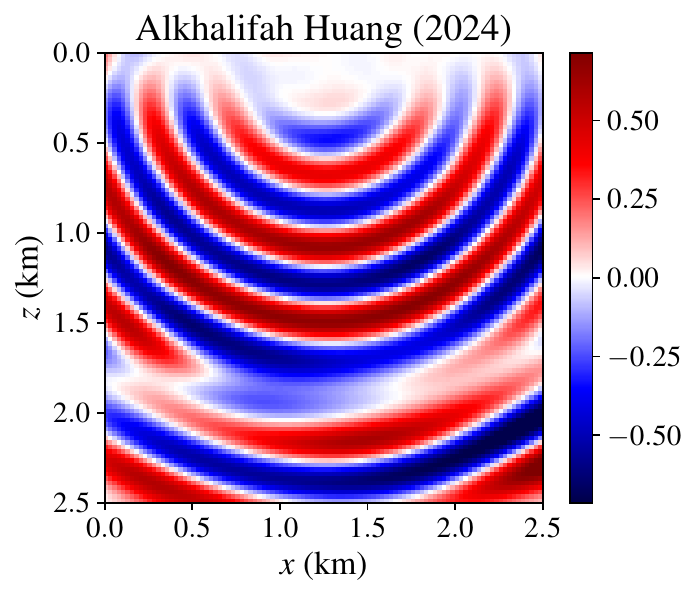}
		%			\caption{Prediction: \cite{Alkhalifah2024} method}
	\end{subfigure}
	\begin{subfigure}[b]{0.32\textwidth}
		\includegraphics[width=\textwidth]{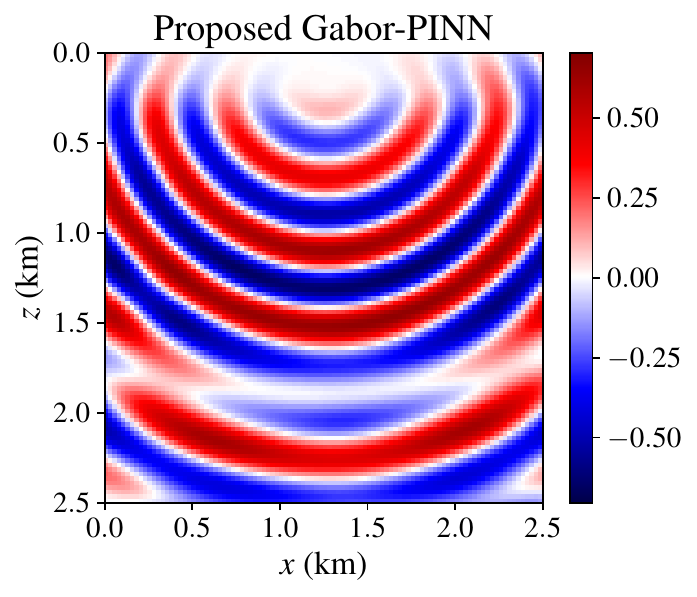}
		%			\caption{Prediction: Proposed Gabor-PINN}
	\end{subfigure}
	\begin{subfigure}[b]{0.32\textwidth}
		\includegraphics[width=\textwidth]{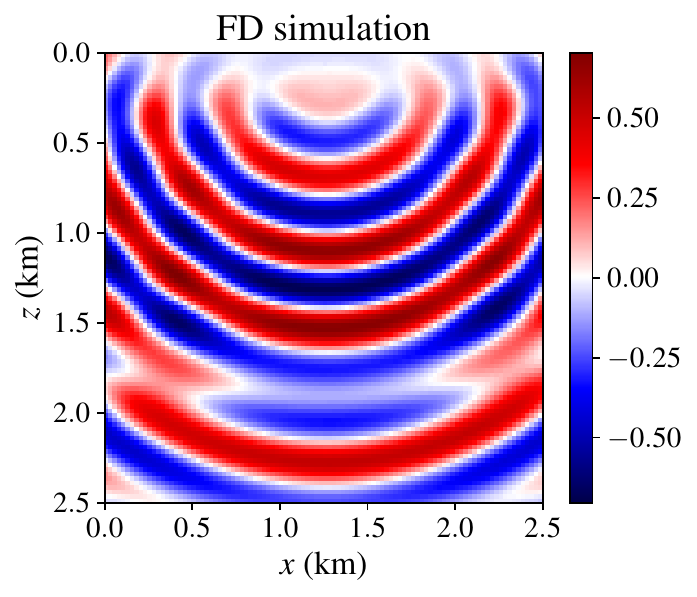}
		%			\caption{Reference wavefield (FD)}
	\end{subfigure}
	
	\caption{Comparison of training losses, validation errors, and scattered wavefield predictions for a 4 Hz wavefield using three different initializations of weights (Test 5). Panels (a) and (b) show the loss curves for \cite{Alkhalifah2024} and our Gabor-PINN, respectively, (c) presents the validation errors for both methods. (d)--(f) display the real part of the wavefield predictions and the finite difference (FD) reference. Our method shows higher accuracy and lower sensitivity to initialization.}
	\label{fig:4Hz_test}
	
\end{figure}

Figure~\ref{fig:4Hz_test} compares the training and validation losses, as well as the validation errors, across three different random initializations of weights. The method of \cite{Alkhalifah2024} shows more sensitivity to initialization, resulting in varying convergence behavior and final errors. In contrast, our proposed Gabor-PINN demonstrates better resilience, with consistently stable loss curves and converging prediction errors. Our method also reaches considerably lower errors, as the plotted mean squared errors (MSE) indicate.

Figures~\ref{fig:4Hz_test} further illustrates the predicted real parts of the wavefields from \cite{Alkhalifah2024}, our proposed method, and a reference wavefield obtained using finite-difference (FD) modeling.

\subsubsection{Test 6: Sensitivity to Model Design}

This test investigates the impact of neural network architecture design on each method's performance and computational efficiency. The same velocity model as in the previous test is used, but with a higher source frequency of 16~Hz. Each method is tested using two architectures: a smaller and a larger network.

Table~\ref{tab:gabor_comparison} summarizes the model characteristics, including the number of neurons per layer, number of Gabor functions, trainable parameters, computational complexity (in MACs), and prediction errors. Our proposed method and the method of \cite{huang2023b} use the same number of layers and neurons, though the latter employs significantly more Gabor functions because they are applied throughout the network. For \cite{Alkhalifah2024}, we use both their original architecture and a modified larger version with a comparable number of trainable parameters to our model. Our method and the \cite{Alkhalifah2024} method use an identical number of Gabor functions and positional encoding with $K=4$. In all methods we use 40,000 collocation points per epoch.

\begin{figure*}[tbp]
	\centering
	\begin{subfigure}[b]{0.32\textwidth}
		\centering
		\includegraphics[width=\textwidth]{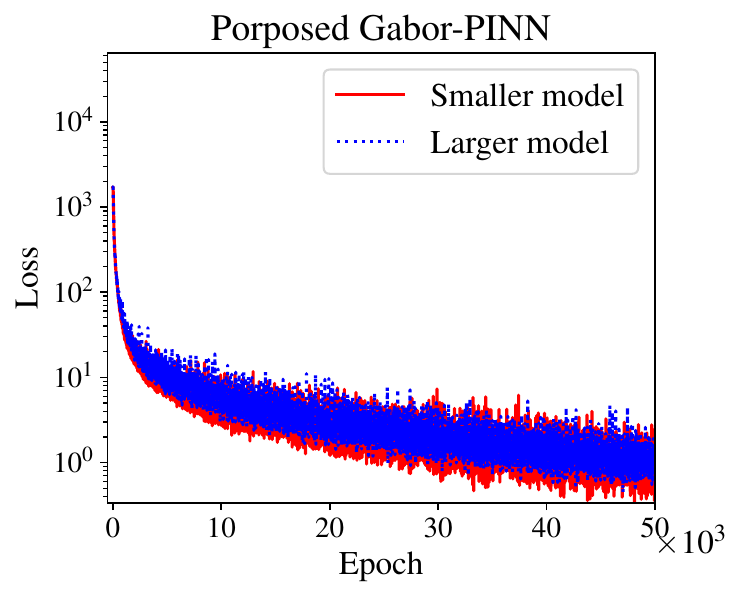}
		
	\end{subfigure}
	\hfill
	\begin{subfigure}[b]{0.32\textwidth}
		\centering
		\includegraphics[width=\textwidth]{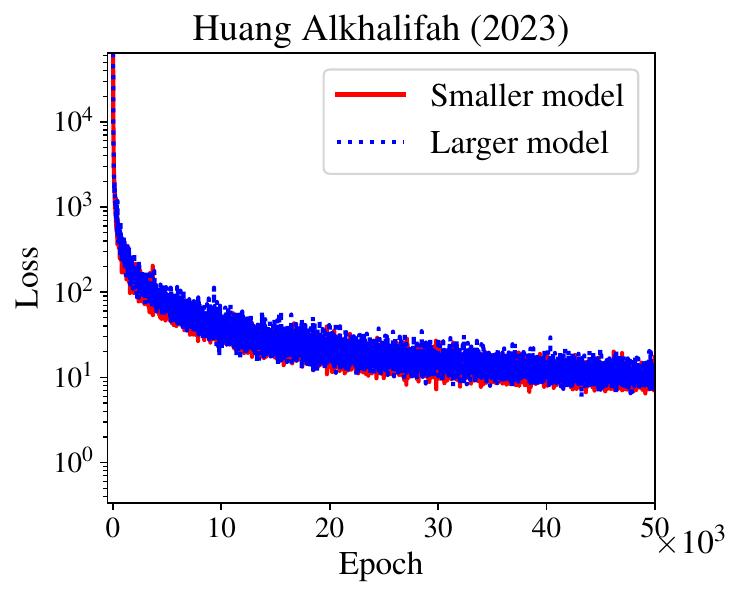}
		
	\end{subfigure}
	\hfill
	\begin{subfigure}[b]{0.32\textwidth}
		\centering
		\includegraphics[width=\textwidth]{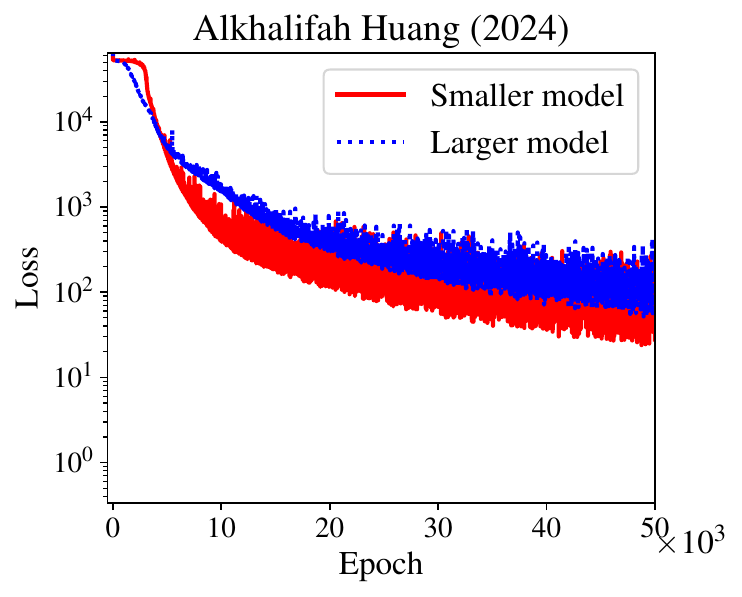}
		
	\end{subfigure}
	
	\begin{subfigure}[b]{0.32\textwidth}
		\centering
		\includegraphics[width=\textwidth]{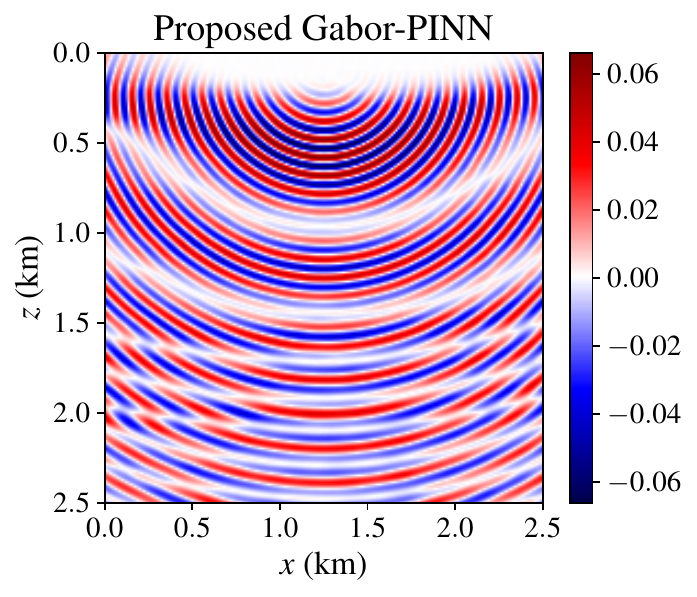}
		
	\end{subfigure}
	\hfill
	\begin{subfigure}[b]{0.32\textwidth}
		\centering
		\includegraphics[width=\textwidth]{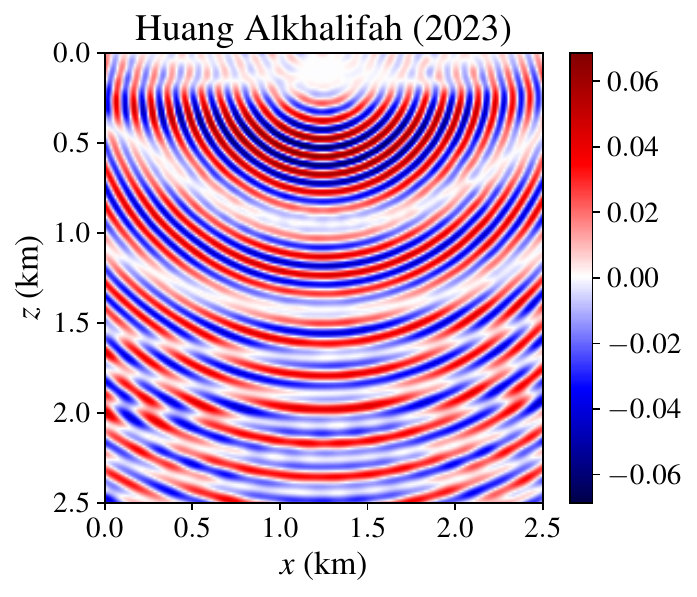}
		
	\end{subfigure}
	\hfill
	\begin{subfigure}[b]{0.32\textwidth}
		\centering
		\includegraphics[width=\textwidth]{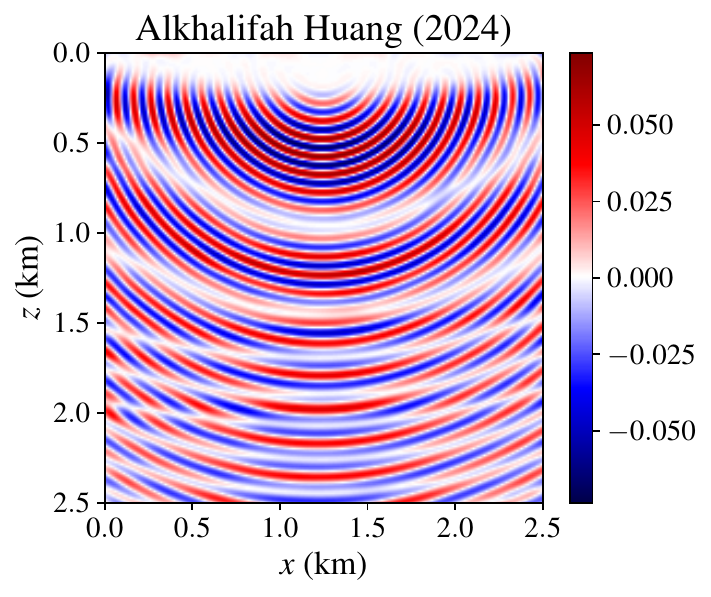}
		
	\end{subfigure}
	
	\begin{subfigure}[b]{0.32\textwidth}
		\centering
		\includegraphics[width=\textwidth]{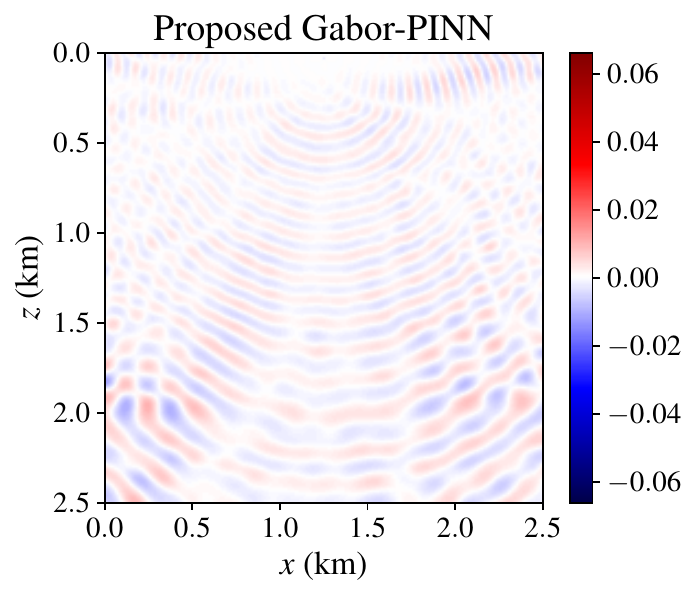}
		
	\end{subfigure}
	\hfill
	\begin{subfigure}[b]{0.32\textwidth}
		\centering
		\includegraphics[width=\textwidth]{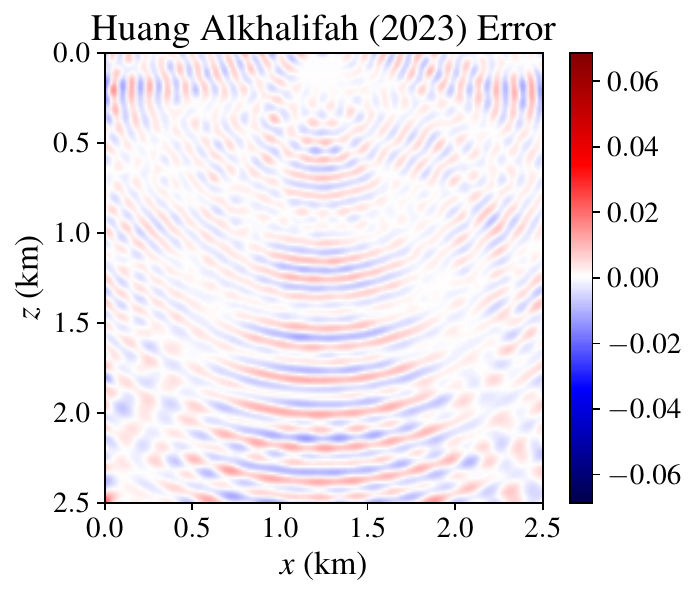}
		
	\end{subfigure}
	\hfill
	\begin{subfigure}[b]{0.32\textwidth}
		\centering
		\includegraphics[width=\textwidth]{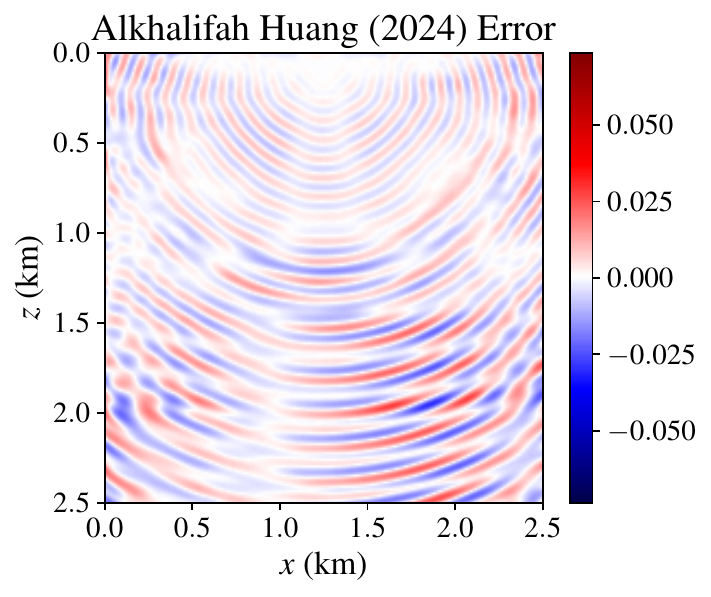}
		
	\end{subfigure}
	\caption{Comparative analysis of the proposed Gabor-PINN, Huang and Alkhalifah (2023), and Alkhalifah and Huang (2024) for a 16~Hz wavefield simulation (Test 6). The first row shows the loss evolution for both small and large models of each method. The second row displays the best predicted wavefields. The third row presents the corresponding prediction errors relative to the reference finite-difference solution. Quantitative results, including network architectures, computational complexity, and final prediction errors, are summarized in Table~\ref{tab:gabor_comparison}.}
	
	\label{fig:comparison_16Hz}
	
\end{figure*}

Figure~\ref{fig:comparison_16Hz} presents comparative results. The top row shows training loss curves for each method under both architectures. The middle row displays the best predicted wavefields, and the bottom row depicts absolute errors relative to an FD reference wavefield. These visualizations offer insights into wavefield quality and spatial accuracy.

Table~\ref{tab:gabor_comparison} also provides a quantitative evaluation of computational complexity. Our proposed Gabor-PINN achieves the lowest MAC count among the tested methods while delivering the most accurate predictions. Moreover, when comparing relative training times measured on different GPU devices, the method of \cite{Alkhalifah2024} takes approximately 1.5 times longer to train than our method, while the method of \cite{huang2023b} takes about 4 times longer. These differences reflect both the added parameter overhead and the broader application of Gabor functions in compared methods.

Importantly, our method demonstrates minimal sensitivity to architectural changes, producing consistent results across both network sizes. In contrast, the accuracy of \cite{Alkhalifah2024} degrades significantly when switching to a different architecture, indicating greater architectural sensitivity. While the method by \cite{huang2023b} performs reasonably well in terms of accuracy, its substantially higher number of parameters and Gabor functions leads to the highest computational burden.

These findings confirm that the proposed Gabor-PINN not only improves predictive accuracy, but also offers greater efficiency and robustness relative to existing Gabor-based PINN methods. This advantage is particularly important for high-frequency problems, where both computational complexity and model sensitivity play a critical role.

\begin{table}[htbp]
	\centering
	\small
	\caption{Comparison of Gabor-PINN models for a 16Hz wavefield simulation (Figure~\ref{fig:comparison_16Hz}): architecture complexity and prediction accuracy after 50{,}000 epochs. 
	\textbf{MACs}: Multiply-Accumulate operations (in thousands), representing computational complexity. 
	\textbf{Params}: Number of trainable parameters, in thousands. 
	\textbf{NMSE}: Normalized Mean Squared Error. 
	\textbf{NMAE}: Normalized Mean Absolute Error. 
	\textbf{Gabor}: Number of Gabor functions used in the network. 
	\textbf{Aux}: Refers to the additional auxiliary network.}
	
	\renewcommand{\arraystretch}{1.1}
	\resizebox{\textwidth}{!}{%
		\begin{tabular}{ l c c c c c c }
			\toprule
			\textbf{Model} & \textbf{Neurons} & \textbf{Gabor} & \textbf{Params} & \textbf{MACs} & \textbf{NMSE} & \textbf{NMAE} \\
			\textbf{} & \textbf{per layer} & \textbf{\#} & \textbf{(k)} & \textbf{(k)} &  &  \\
			\midrule
			\multicolumn{1}{l}{\shortstack[l]{\textbf{Our proposed} \\ Smaller model}} & 
			128-128-128 & 64 & 36.22 & 36.42 & 0.0107 & 0.0990 \\
			\multicolumn{1}{l}{\shortstack[l]{\textbf{} \\ Larger model}} & 
			128-128-128-128 & 64 & 52.74 & 52.87 & 0.0102 & 0.1004 \\
			\midrule
			\multicolumn{1}{l}{\shortstack[l]{\textbf{\cite{huang2023b}} \\ Smaller model}} & 
			128-128-128 & 512 & 53.89 & 51.84 & 0.0203 & 0.1377 \\
			\multicolumn{1}{l}{\shortstack[l]{\textbf{} \\ Larger model}} & 
			128-128-128-128 & 640 & 71.43 & 68.87 & 0.0179 & 0.1304 \\
			\midrule
			\multicolumn{1}{l}{\shortstack[l]{\textbf{\cite{Alkhalifah2024}} \\ Smaller model}} & 
			128-128-64-64-64+Aux & 64 & 35.82 & 39.74 & 0.0722 & 0.2580 \\
			\multicolumn{1}{l}{\shortstack[l]{\textbf{} \\ Larger model}} & 
			128-128-128-96-64+Aux & 64 & 54.35 & 60.31 & 0.3253 & 0.5284 \\
			\bottomrule
		\end{tabular}
	}
	\label{tab:gabor_comparison}
\end{table}

\section*{Conclusion}

We proposed a Physics-Informed Neural Network (PINN) framework that incorporates explicit, trainable Gabor basis functions for simulating scattered wavefields governed by the Helmholtz equation. Unlike prior approaches that used Gabor functions as multiplicative activations throughout the network or relied on auxiliary networks to learn both the Gabor centers and magnitudes, our method substantially reduces architectural complexity. This is achieved by redefining the network's task as learning a mapping from input coordinates to a custom Gabor coordinate system. Since Gabor functions inherently encode the oscillatory behavior of wavefields, the learned mapping can be smooth and easier to optimize. This formulation enables a simplified implementation of Gabor basis functions by absorbing the effects of both the Gabor centers and magnitudes into a single coordinate transformation.

The proposed incorporation of Gabor functions maintains the total number of trainable weights by balancing the new Gabor-specific parameters with a reduced number of weights in the final layer. As a result, our method introduces no net increase in model size. Compared to the method of Huang and Alkhalifah (2023), our approach significantly reduces the number of Gabor functions, trainable parameters, and computational complexity, while also eliminating the need for a manually tuned frequency scaling factor. Relative to Alkhalifah and Huang (2024), we avoid the instability of multi-network training and remove the need to select the size of an auxiliary network---another influential hyperparameter. Numerical experiments demonstrate that our simplified Gabor-PINN provides key benefits over previous Gabor-based PINN implementations: higher accuracy, improved stability, robustness to architectural choices and initialization, and reduced training cost.

We also introduced an efficient implementation of the Perfectly Matched Layer (PML) for the scattered Helmholtz equation by deriving a real-valued loss formulation, which avoids the need for complex-valued training, and by proposing an analytical expression for the background wavefield. This improves the boundary behavior and accuracy with no additional computational cost, apart from the inevitable increase in domain size due to PML inclusion.

The proposed Gabor-PINN framework is expected to generalize well to other differential equations, particularly those with oscillatory solutions. However, equation-specific elements---such as loss functions---would need to be adapted to the structure of each PDE. While our method simplifies the training task and architecture, further simplification may be possible by adopting plane wave basis functions, which will be investigated in future work. We also plan to explore hybrid optimization strategies to mitigate residual low-frequency bias and enhance the resolution of finer wavefield details.

\section*{CRediT authorship contribution statement}

\textbf{Mohammad Mahdi Abedi:} Conceptualization, Methodology, Writing – original draft, Data curation, Investigation, Software, Visualization, Writing – review \& editing.  
\textbf{David Pardo:} Conceptualization, Funding acquisition, Supervision, Writing – review \& editing.  
\textbf{Tariq Alkhalifah:} Conceptualization, Supervision, Writing – review \& editing.

\section*{Declaration of competing interest}
The authors declare that they have no known competing financial interests or personal relationships that could have appeared to influence the work reported in this paper.

\section*{Code Availability}  
The implementation of the proposed Gabor-Enhanced Physics-Informed Neural Network (PINN) for solving the Helmholtz equation is available as an open-source repository. The code, along with instructions for usage and reproducibility, can be accessed at:  
\begin{center}
	\url{https://github.com/mahdiabedi/Gabor-Enhanced-PINN}
\end{center}  

\section*{Acknowledgments}
This research was supported by the following Research Projects/Grants: European Union's Horizon Europe research and innovation programme under the Marie Sklodowska-Curie grant agreement No 101119556; TED2021-132783B-I00 funded by MCIU/AEI/10.13039/501100011033 and by the European Union NextGenerationEU/PRTR; PID2023-146678OB-I00 funded by MCIU/AEI/10.13039/501100011033 and by FEDER, UE; “BCAM Severo Ochoa” accreditation of excellence CEX2021-001142-S funded by MICIU/AEI/10.13039/501100011033; Basque Government through the BERC 2022-2025 program; BEREZ-IA (KK-2023/00012) and RUL-ET (KK-2024/00086), funded by the Basque Government through ELKARTEK; Consolidated Research Group MATHMODE (IT1456-22) given by the Department of Education of the Basque Government; BCAM-IKUR-UPV/EHU, funded by the Basque Government IKUR Strategy and by the European Union NextGenerationEU/PRTR.

\appendix
\section{Scattered Wave Equation with PML}
\label{app:PML}

	\renewcommand{\theequation}{A.\arabic{equation}}
	\setcounter{equation}{0} % Reset equation counter for appendix
	
	 For an isotropic medium with wave velocity $v(\mathbf{x})$, the Helmholtz equation with PML is given by \citep{berenger1994perfectly,chen2013optimal,song2022versatile}:
	\begin{equation}
		\frac{\partial}{\partial x}\left(\frac{e_z}{e_x} \frac{\partial u}{\partial x} \right)+ \frac{\partial}{\partial z}\left(\frac{e_x}{e_z} \frac{\partial u}{\partial z} \right)+ {e_z}{e_x} \omega^2 m(\mathbf{x}) u(\mathbf{x}) = f(\mathbf{x}),
		\label{eq:helmholtz_pml}
	\end{equation}
	where $m(\mathbf{x}) = \frac{1}{v^2(\mathbf{x})}$ is the slowness squared. The terms $e_x$ and $e_z$ represent the complex stretching along the $x$ and $z$ directions, respectively, in the PML regions, defined in Equation \ref{eq:ex_ez}. The solution $u(\mathbf{x}) = u^r(\mathbf{x}) + \mathrm{i} u^i(\mathbf{x})$ is a complex-valued wavefield, where $\mathbf{x} = \{x, z\}$ denotes the spatial coordinates in a 2D domain.
		
		To derive the scattered Helmholtz equation, we decompose the total wavefield into a background wavefield \( u_0(\mathbf{x}) \) and a scattered wavefield \( u_s(\mathbf{x}) \):
		
		\begin{equation*}
			u(x) = u_0(\mathbf{x}) + u_s(\mathbf{x}), 
		\end{equation*}
and substitute this into Equation \eqref{eq:helmholtz_pml}, to obtain:
		
		\begin{equation}
			\frac{\partial}{\partial x} \left( \frac{e_z}{e_x} \frac{\partial (u_0 + u_s)}{\partial x} \right) +
			\frac{\partial}{\partial z} \left( \frac{e_x}{e_z} \frac{\partial (u_0 + u_s)}{\partial z} \right) +
			e_x e_z \omega^2 m(x) (u_0 + u_s) = f(x).
			\label{eq:A8}
		\end{equation}
	The background wavefield \( u_0(x) \) should satisfy the Helmholtz equation with PML \eqref{eq:helmholtz_pml} in a homogeneous medium:
		
		\begin{equation}
			\frac{\partial}{\partial x} \left( \frac{e_z}{e_x} \frac{\partial u_0}{\partial x} \right) +
			\frac{\partial}{\partial z} \left( \frac{e_x}{e_z} \frac{\partial u_0}{\partial z} \right) +
			e_x e_z \omega^2 m_0 u_0 = f(x),
				\label{eq:A9}
		\end{equation}
where \( m_0 = \frac{1}{v_0^2} \) is the background slowness squared.
		
The amplitude decay inside the PML in $x$ direction is given by:
\begin{equation}
	e^{-\int \frac{\omega c l_x^2}{v_0 (1 + c^2 l_x^4)} \, dx}.
\end{equation}
Using the analytical solution for the background wavefield in a homogeneous medium and approximating the amplitude decay for small $c$, we find the analytical approximation of $u_0(\mathbf{x})$ that satisfies \eqref{eq:A9}:

	\begin{equation}
u_0(\mathbf{x}) = \frac{i}{4} H_0^{(2)}\left(\frac{\omega}{v_0}  |\mathbf{x} - \mathbf{x}_f|\right) e^{-\omega c \frac{(l_x^2+l_z^2)}{3 v_0}^{(3/2)}},
	\end{equation}
where $H_0^{(2)}\left(.\right)$ is the zero-order Hankel function of the second kind.

Subtracting  \eqref{eq:A9}  from the total wave equation  \eqref{eq:A8} results in the scattered Helmholtz equation with PML:
		
		\begin{equation}
			\frac{\partial}{\partial x} \left( \frac{e_z}{e_x} \frac{\partial u_s}{\partial x} \right) +
			\frac{\partial}{\partial z} \left( \frac{e_x}{e_z} \frac{\partial u_s}{\partial z} \right) +
			e_x e_z \omega^2 m(x) u_s = - e_x e_z \omega^2 \delta m(x) u_0(x).
			\label{eq:scattered_helmholtz_pml}
		\end{equation}

	Using the mean squared difference between the two sides of the above equation over $N$ collocation points to define loss function $\mathcal{L}$, we obtain:
	\begin{equation}
		\mathcal{L} = \frac{1}{N} \sum \left( {L}_r^2 + {L}_i^2 \right),
	\end{equation}
	where ${L}_r$ and ${L}_i$ are the real and imaginary parts of the loss. For implementation efficiency, we explicitly calculate the real and imaginary parts of the loss to avoid using complex numbers during training. 
	
	The terms ${L}_r$ and ${L}_i$ are calculated as:
	\begin{align}
		{L}_r &= F_{1}^r + F_{2}^r +\left( 1 - c^2 l_x^2 l_z^2 \right) \omega^2 \left( m  u_s^r +   u_0^r \delta m \right) 
		+ c \omega^2 \left( l_x^2 + l_z^2 \right) \left( m  u_s^i +   u_0^i \delta m\right), \\
		{L}_i &= F_{1}^i + F_{2}^i  +\left( 1 - c^2 l_x^2 l_z^2 \right) \omega^2 \left( m  u_s^i +  u_0^i \delta m\right) 
		- c \omega^2 \left( l_x^2 + l_z^2 \right) \left( m  u_s^r +   u_0^r \delta m\right),
	\end{align}
using the following definitions for the auxiliary terms,
\begin{equation}
	\begin{aligned}
		C_1 &= \frac{1 + c^2 l_x^2 l_z^2}{1 + c^2 l_x^4}, \quad
		C_2 = \frac{c (l_x^2 - l_z^2)}{1 + c^2 l_x^4}, \\
		C_3 &= \frac{1 + c^2 l_x^2 l_z^2}{1 + c^2 l_z^4}, \quad
		C_4 = \frac{c ( l_z^2-l_x^2 )}{1 + c^2 l_z^4}.
	\end{aligned}
\end{equation}
	and calculating the derivatives of these auxiliary terms,
	\begin{equation}
	\begin{aligned}
		\frac{\partial C_1}{\partial x} &= \frac{2 c^2 l_x \big(-2 l_x^2 + l_z^2 - c^2 l_x^4 l_z^2 \big) }{\big(1 + c^2 l_x^4\big)^2}\frac{\partial l_x}{\partial x}, \\
		\frac{\partial C_2}{\partial x} &= \frac{2 c l_x \big(1 - c^2 l_x^2 (l_x^2 - 2 l_z^2)\big) }{\big(1 + c^2 l_x^4\big)^2}\frac{\partial l_x}{\partial x}, \\
		\frac{\partial C_3}{\partial z} &= \frac{2 c^2 l_z \big(-2 l_z^2 + l_x^2  - c^2 l_z^4 l_x^2\big) }{\big(1 + c^2 l_z^4\big)^2}\frac{\partial l_z}{\partial z}, \\
		\frac{\partial C_4}{\partial z} &= \frac{2 c l_z \big(1 - c^2 l_z^2 (-2 l_x^2 + l_z^2)\big) }{\big(1 + c^2 l_z^4\big)^2}\frac{\partial l_z}{\partial z}.
	\end{aligned}
	\end{equation}
	we find:
	\begin{equation}
	\begin{aligned}
		F_{1}^r &= \frac{\partial C_1}{\partial x} \frac{\partial u_s^r}{\partial x} 
		+ C_1 \frac{\partial^2 u_s^r}{\partial x^2}
		- \frac{\partial C_2}{\partial x} \frac{\partial u_s^i}{\partial x}
		- C_2 \frac{\partial^2 u_s^i}{\partial x^2}, \\
		F_{2}^r &= \frac{\partial C_3}{\partial z} \frac{\partial u_s^r}{\partial z} 
		+ C_3 \frac{\partial^2 u_s^r}{\partial z^2}
		- \frac{\partial C_4}{\partial z} \frac{\partial u_s^i}{\partial z}
		- C_4 \frac{\partial^2 u_s^i}{\partial z^2}, \\
		F_{1}^i &= \frac{\partial C_1}{\partial x} \frac{\partial u_s^i}{\partial x} 
		+ C_1 \frac{\partial^2 u_s^i}{\partial x^2}
		+ \frac{\partial C_2}{\partial x} \frac{\partial u_s^r}{\partial x}
		+ C_2 \frac{\partial^2 u_s^r}{\partial x^2}, \\
		F_{2} ^i &= \frac{\partial C_3}{\partial z} \frac{\partial u_s^i}{\partial z} 
		+ C_3 \frac{\partial^2 u_s^i}{\partial z^2}
		+ \frac{\partial C_4}{\partial z} \frac{\partial u_s^r}{\partial z}
		+ C_4 \frac{\partial^2 u_s^r}{\partial z^2}.
	\end{aligned}
\end{equation}

	From the above equations, we can see that with the addition of PML, each term of the loss depends on both the real and imaginary components of the NN output. The first-order derivatives of $l$ and one extra hyper-parameter $a_0$ also contribute to the added complexity.   

\section*{References}
\bibliographystyle{apalike}
\bibliography{references}

\end{document}